Intelligent Systems Group
University of Siegen

# Dataset Pruning in RecSys and ML: Best Practice or Mal-Practice?

Bachelor's Thesis
*Computer Science*
*Leonie Winter*
*1097962*

30.09.2025

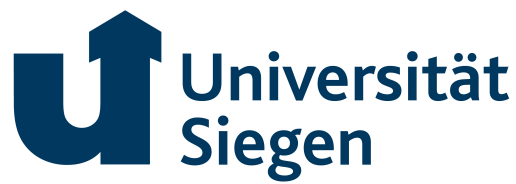

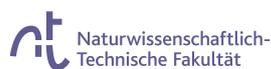 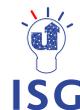

**Examiners**
Prof. Dr. Joeran Beel — Intelligent Systems Group
M. Sc. Tobias Vente — Intelligent Systems Group

This Bachelor's Thesis is handed in according to the requirements of the University of Siegen for the study program Bachelor Computer Science of the year 2012 (PO 2012).

**Process period**
12.05.2025 to 30.09.2025

**Examiners**
Prof. Dr. Joeran BEEL
M. Sc. Tobias VENTE

## **Eigenständigkeitserklärung**

Ich versichere, dass ich die schriftliche Ausarbeitung selbständig angefertigt und keine anderen als die angegebenen Hilfsmittel benutzt habe. Alle Stellen, die dem Wortlaut oder dem Sinn nach (inkl. Übersetzungen) anderen Werken entnommen sind, habe ich in jedem einzelnen Fall unter genauer Angabe der Quelle (einschließlich des World Wide Web sowie generativer KI und anderer elektronischer Datensammlungen) deutlich als Entlehnung kenntlich gemacht. Dies gilt auch für angefügte Zeichnungen, bildliche Darstellungen, Skizzen und dergleichen. Ich nehme zur Kenntnis, dass die nachgewiesene Unterlassung der Herkunftsangabe als versuchte Täuschung gewertet wird.

Ich bestätige, dass die elektronische Version inhaltlich mit der gedruckten Version übereinstimmt.

______________________  ______________________________________

Datum  Unterschrift

*Declaration of Authorship*

*I certify that I have prepared this written work independently and have not used any aids other than those specified. I have clearly identified all passages that are taken from other works in terms of wording or meaning (including translations) as borrowed material in each individual case, stating the exact source (including the World Wide Web as well as generative AI and other electronic data collections). This also applies to attached drawings, pictorial representations, sketches and the like. I acknowledge that the proven omission of the indication of origin will be regarded as attempted deception.*

*I confirm that the content of the electronic version is the same as the printed version.*

# I. Contents

## II. Abstract

Offline evaluations in recommender system research depend heavily on datasets, many of which are pruned, such as the widely used MovieLens collections. This thesis examines the impact of data pruning – specifically, removing users with fewer than a specified number of interactions – on both dataset characteristics and algorithm performance. Five benchmark datasets were analysed in their unpruned form and at five successive pruning levels (5, 10, 20, 50, 100). For each coreset, we examined structural and distributional characteristics and trained and tested eleven representative algorithms. To further assess if pruned datasets lead to artificially inflated performance results, we also evaluated models trained on the pruned train sets but tested on unpruned data.

Results show that commonly applied core pruning can be highly selective, leaving as little as 2% of the original users in some datasets. Traditional algorithms achieved higher nDCG@10 scores when both training and testing on pruned data; however, this advantage largely disappeared when evaluated on unpruned test sets. Across all algorithms, performance declined with increasing pruning levels when tested on unpruned data, highlighting the impact of dataset reduction on the performance of recommender algorithms.

# III.  Acknowledgements

I would like to express my gratitude to Prof. Dr Ing. Beel, as well as to Lukas Wegmeth and Tobias Vente, for their supervision and valuable guidance and insight throughout the creation of this thesis.

Throughout the course of writing this thesis, ChatGPT was used as an auxiliary writing tool, particularly for phrasing suggestions and proofreading. All suggestions from ChatGPT were carefully reviewed, edited and adapted to ensure correctness and consistency with my own voice. This practice aligns with the guideline that generative models can be used to support writing, but their use should be disclosed transparently. [7]

The code for all experiments conducted during this thesis is openly accessible at GitHub [51].

# 1. Introduction

## 1.1 Background

The performance of recommender systems is typically assessed through offline evaluations, online evaluations, or user studies. Offline evaluations – training and testing algorithms on datasets – are the most common first step to narrow down promising algorithms. [14, 41, 54] Ideally, the datasets for offline evaluations should match the intended application domain [15, 25]; nonetheless, many studies rely on a small set of benchmark datasets, some of which have been pruned, meaning specific interactions have been removed [6, 8, 26, 27, 39, 47].

Data pruning has been shown to alter structural characteristics of datasets, which are key factors in explaining performance variance across datasets (e.g. density) [1, 15, 18, 20]. A commonly applied pruning technique is the systematic removal of low-activity users from datasets [20, 54, 55]. These so-called cold-start users are a recognised challenge in recommender-system research, and removing them from both the training and test sets effectively avoids evaluating recommender performance on this group [20, 24, 30]. This can lead to artificially inflated performance scores [8] and limit generalizability [6, 15, 20], as pruning causes the test data to diverge from real-world scenarios, resulting in datasets composed primarily of high-activity users with rich interaction histories [20].

As more complex models emerge and the demand for sustainable recommender systems grows, interest in data reduction techniques such as data pruning has increased [5, 9, 43, 45, 49, 58], even though data pruning is recognised as a barrier to comparability and reproducibility in recommender-system research [17, 22, 27, 47, 54]. These contrasting developments warrant closer examination of how data pruning affects both dataset characteristics and algorithm performance.

## 1.2 Research Problem

Data pruning has been established as a popular step in dataset preprocessing, as demonstrated in reviews. Sun et al. found that of more than 42 examined papers that explicitly stated preprocessing datasets, 60% applied Data Pruning in this process [47]. This aligns with Beel and Brunel[8], who found that 65% of the papers they examined that employed offline evaluations used either a pruned dataset or pruned a dataset during the preprocessing steps.

While data pruning has been proven to positively impact runtime, computational load, and dataset density, seemingly achieving improved performance values [4], a good result in performance evaluation on a pruned dataset is not generally transferable to other datasets.

In recommender-system research, so-called cold-start users or items present a well-known challenge, as many algorithms struggle to generate reliable recommendations for users without a rich interaction history. Newly introduced items lacking similarity to existing ones pose the same challenge. [2, 29, 30, 42]

These challenging user groups are effectively eliminated when user-based data pruning is applied. As a result, algorithm evaluations on pruned datasets are biased toward high-activity users, since the challenging low-activity group is progressively excluded from both training and testing. [24] In practice, applying the MovieLens-standard 20-core pruning can exclude up to 42% of users while still retaining about 95% of interactions [8].

Algorithm performance partly depends on the dataset structure. For instance, density is a key factor in explaining performance variance. [14, 16] Therefore, it is possible to influence algorithm performance through the deliberate selection of datasets or coresets. However, higher levels of pruning cause a greater divergence from the raw data that a recommender is likely to encounter in real-world scenarios. [20] Consequently, pruned datasets can hinder generalisation and comparability when evaluating algorithms.

Although data pruning is a commonly applied procedure, 44% of researchers still do not report the preprocessing applied to the datasets used in their studies, as a recent review showed [55]. This lack of transparency is problematic because it greatly hinders the reproducibility of algorithm performance and the ability to make fair comparisons.

In summary, while dataset pruning is a widely acknowledged tool, it poses three key problems: reproducibility suffers when pruning is undocumented, comparability is compromised when coresets are deliberately selected, and generalisability declines as pruned datasets diverge from real-world data.

Despite growing awareness of these issues, the combined effects of progressive data pruning on dataset characteristics and algorithm performance remain insufficiently examined. The lack of systematic analysis of how progressive data pruning alters dataset characteristics and affects the measured performance and ranking of recommender algorithms creates uncertainty in evaluating and comparing recommendation models. Addressing this gap is essential for reproducibility, fairness, and the reliable development of recommender systems.

## 1.3 Research Question

Building on the previous section, this thesis aims to mitigate the gap identified in research on data pruning and its impact on algorithm performance and the structural characteristics of datasets. Unlike previous studies, we will investigate the changes in distributional characteristics of the pruned datasets, with a focus on user, item, and interaction retention. By progressively increasing the pruning core-level of datasets, comparing the performance of 11 algorithms on these pruned variations of the datasets and examining the changes in dataset characteristics, we aim to address the following research question:

*How does progressive dataset pruning by increasing core thresholds affect the performance of recommender systems, the structural and distributional characteristics of the dataset?*

## 1.4 Research Objective

The objective of this thesis is two-fold. First, we aim to assess how progressive pruning affects dataset characteristics and to quantify the proportion of the original data that remains at each pruning level. Second, we strive to determine whether progressive pruning alters the ranking of algorithms and whether it artificially inflates their performance scores.

# 2. Background

Recommender systems have become indispensable for delivering personalised recommendations across diverse application areas, including e-commerce, streaming services, social networks, and education [29, 31, 35, 42]. The development of recommender systems requires both models and datasets, yet it is the datasets on which different models are trained, evaluated and compared in offline studies. Consequently, datasets play a foundational role in recommender-system research, and selecting the right dataset is a critical step in the development process [47]. Some datasets have become benchmarks, widely used for training and evaluation in recommender system research. For example, the MovieLens datasets [53], which a recent review found to be used in approximately 50% of studies [2].

The most commonly used variants of the MovieLens datasets are pruned. The term data pruning refers to setting a threshold *t* for a group (e.g., users or items) and eliminating all interactions from any user or item whose total interactions fall below that threshold [20]. A well-known example is the MovieLens datasets (except ML Full Latest), which apply *t = 20* to users, retaining only interactions from users with at least 20 interactions in the original dataset. Such a subset is called 20-core pruned, where core denotes the minimum interaction count required for inclusion. In this thesis, 0-core refers to the full, unpruned dataset, which has only been randomly downsampled.

Although data pruning is sometimes also referred to as data filtering, it differs fundamentally from *downsampling*, in which a subset is drawn (e.g., randomly) without intentionally altering the dataset's inherent characteristics [43]. In contrast, subjecting datasets to pruning methods alters the structural characteristics of the datasets, such as the average number of ratings per user or item, or the density of the dataset, which describes the fraction of ratings in the dataset relative to all possible user-item interactions [1, 18]. These characteristics are part of our analysis and are therefore briefly outlined in the *Methodology* chapter.

## 3. Related Work

The question of how to create sound evaluation for recommender systems [6, 54] and machine learning algorithms is an active field of research that spans multiple factors, including the impact of reduction strategies (e.g., dataset pruning) on dataset composition and algorithm performance [13, 34, 58]. Various strategies aim to decrease the computational load of evaluations or improve model performance and efficiency by altering the datasets or models, including model pruning (e.g., by removing weights or nodes) [11, 53], downsampling, and the intentional removal of targeted interactions. Removing interactions from a dataset to alter its composition or quality can take several forms, such as data pruning, noise reduction (e.g., via sampling techniques) [28], or the removal of users with a detrimental impact on performance [36].

To the best of our knowledge, Beel and Brunel [8] were the first to review a set of conference papers on dataset usage to determine the prevalence of pruned datasets in recommender-system evaluations. Their review found that almost 50% of offline evaluations were conducted on at least partially pruned datasets, with the 20-core pruned MovieLens datasets being the most frequently used. In assessing the impact of 20-core pruning on algorithm performance, they compared six collaborative algorithms on an unpruned version of MovieLens with a version containing only users who exceeded the 20-interaction threshold. Their results show increased performance on the pruned dataset; however, it is worth noting that the test sets were also pruned. While this paper provides the foundation for our work, its performance comparison was limited to a single dataset and reported only error-based metrics (RMSE, MAE). In this thesis, we extend the analysis to ranking-based metrics and evaluate a broader variety of algorithms across a wider range of datasets and pruning levels.

A recent review on standard practices in recommender evaluation has found that dataset selection and preprocessing techniques, including pruning, vary widely across the field [55]. Zhao et al. found that 44% of studies did not report any preprocessing details and 34% implicitly applied n-core filtering with thresholds of 5 or 10. Thresholds varied widely, with some papers using 25- or 30-core pruning, but the authors' own experiments focused only on 5- and 10-core filters. They observed that these two thresholds produced significantly different rankings on some datasets (e.g., the sparse Amazon datasets). Although they reported the proportion of users excluded, exploring a broader range of core thresholds and conducting a more thorough analysis of changes in the datasets' structural and distributional characteristics would be valuable.

Sun et al. [47] had earlier taken a similar approach, finding that 60% of studies applied 5- or 10-core filtering, with some using thresholds as high as 20 or 30. The authors also examined the 5- and 10-core filters, which Zhao et al. later confirmed to be aligned with.

Bentzer and Thulin [10] investigated a user-based approach of data pruning by limiting the number of users included in their training sets. The base dataset was MovieLens-25M, already 20-core pruned for users, and it was further filtered to retain only items that had been interacted with by at least 10% of users. The authors then evaluated SVD and IBCF on progressively smaller test sets and found IBCF to achieve higher accuracy on small datasets, whereas SVD was preferable for larger datasets, despite a slight loss in accuracy, because of its lower execution time. Their results highlight that, in certain scenarios, accepting a slight decrease in performance can be a worthwhile trade-off for faster execution. However, because their experiment only included two algorithms, the findings have limited generalisability.

Driven by the goal of developing more sustainable and efficient recommender systems, data pruning has also become a focus in studies on data-reduction techniques. Spillo et al. [45] investigated the trade-offs between data reduction and algorithm performance, as well as the impact on the carbon footprint of recommender systems. In their study, the authors reduced the data only after splitting the datasets into training and test sets, thereby manipulating only the size of the training data. They confirmed that data reduction decreases both emissions and performance, while some algorithms showed scenario-dependent potential for achieving acceptable trade-offs.

Similarly, Arabzadeh et al. [4, 5] examined whether a trade-off between energy efficiency and accuracy is possible by applying two user-centred downsampling approaches to datasets. Unlike Spillo et al., they also incorporated core pruning of both users and items at levels 10 and 30, which left only a fraction of the datasets viable for evaluation. Their results showed that more data generally had a positive impact on accuracy, but aligned with Spillo et al., that scenario-dependent trade-offs can be found for individual algorithms. While this study makes a valuable contribution to more sustainable recommender systems, the analysis of dataset characteristics did not extend beyond a basic examination of structural properties.

Regarding dataset characteristics, Deldjoo et al. [18] proposed a rich framework of structural and distributional properties, which we partially adopt in our analysis. Their focus was on assessing the impact of individual dataset characteristics and identifying those with the most significant influence. Shape, Interactions per User or Item, and Density have emerged as impactful characteristics in their results, showing that deliberate dataset selection, focusing on key characteristics, can directly impact performance. In this thesis, we adopt some of the proposed explanatory variables of their study and examine their evolution across increasingly pruned coresets.

Fan et al. [24] explored the MovieLens datasets by selectively removing interactions per user (e.g., the first 15) to assess which interactions are most informative for preference elicitation. They confirmed that MovieLens is a statistical outlier in its distributional characteristics. Their results also showed that, because users were initially encouraged to rate popular items until a threshold was reached during the collection of interactions for the MovieLens datasets, these early 15 interactions contributed little to shaping user preferences.

Further research on the influence of the datasets on algorithm performance was also conducted by Chin et al. [15]; however, their dataset selection did not include pruned variations. Doerfel et al. [20] examined how dataset pruning affects the performance ranking of recommender systems in the context of social bookmarking. They found that density increased with higher core levels, but the low-core settings of 2 and 3 were most consistent with the raw data.

# 4. Methodology

## 4.1 Datasets

This subsection outlines the dataset selection, all preprocessing steps and the train-test splitting strategy applied. Additionally, we outline the key characteristics in our dataset analysis.

### 4.1.1 Dataset Selection

The selection of datasets is based on several reviews, identifying the most commonly used datasets in recommender-system research. The popularity of the MovieLens datasets is undisputed, with a range of 40% [8], 50% [2], and up to 84% [6] of studied papers employing at least one version of MovieLens in their experiments [12, 17, 25, 26, 39, 47, 15]. There are multiple versions of the MovieLens dataset available, with only the MovieLens Full Latest being unpruned and usable for our experiment. [27]

In recommender research, the Amazon [38] datasets are also widely adopted, with 35% [8] to 42% [6] of examined papers including at least one version in their experimentation [25, 39, 47].

Another benchmark dataset in recommender-systems research is the Yelp dataset[59], which is employed in multiple versions from different timeframes [47, 17, 26, 15].

The aforementioned datasets are all classed as explicit datasets, expressing users' preferences for items by numerical ratings ranging from 1 (lowest) to 5 (highest). To employ an implicit dataset in our experiment as well, we chose the Gowalla [16] dataset, which consists of point-of-interest data where every interaction with an item is recognised as a positive preference value, and all non-interacted items are considered negative [25].

For readability, shortened dataset names are used throughout this thesis. The Amazon Toys & Games and Amazon CD & Vinyl datasets are referred to as Amazon Toys and Amazon CD, respectively. Similarly, all references to MovieLens denote the MovieLens Full Latest dataset unless otherwise specified.

As the analysis of structural and distributional characteristics of the datasets is part of this study, please refer to Chapter 5.1 for detailed statistics of each pruned version and the non-pre-processed datasets.

### 4.1.2 Preprocessing of Datasets

All datasets selected for the experiment have been submitted to the same procedure of preprocessing, consisting of transformation into an implicit dataset for originally explicit datasets, removal of duplicate interactions, downsampling to a random subset of fixed size, pruning to six different core-subsets, data splitting and the creation of atomic files. The specific actions taken in each of these steps are listed in detail below, in the same sequence as they were executed.

**Transforming explicit datasets to implicit datasets** All datasets, except for the inherently implicit Gowalla dataset, have been transformed into implicit datasets by retaining all interactions with rating values equal to or greater than 4.0 [3, 44]. These ratings were assigned the binary value of 1.

**Downsampling** For all implicit datasets with more than 3,000,000 interactions, a random subset has been extracted using Pandas' sample function on the interaction dataframe. The resulting subset had a length of 3,000,000 for each dataset. Hereinafter, this downsampled dataset will be referred to as the unpruned or 0-core dataset.

**Coreset Creation** Each dataset was retained in an unpruned version, referred to as 0-core, which retained all interactions by all users and with all items included in the downsampled set. Subsequently, the dataset was pruned in five additional coresets, where each core represented the minimum count of interactions a user had to have in the dataset to be retained in the respective coreset. For example, the 5-coreset retained all interactions by users with at least five interactions in the dataset. Each of the selected datasets was pruned into 5-core, 10-core, 20-core, 50-core, and 100-core subsets.

Please note that the MovieLens datasets are commonly described as 20-core pruned [8], but in practice, the threshold of 20 is only applied to users [27]. Items remain in the dataset even if they have been interacted with only once. Strictly speaking, the term core filtering would be more accurate, since core pruning can imply a recursive process that continues removing interactions until both users and items meet the minimum interaction threshold. For clarity, in this thesis, we use the term '20-core pruned' to denote datasets filtered by a 20-interaction user threshold, which may still include items with as few as one interaction.

**Data Splitting**. For each core subset and the unpruned variant of each dataset, a user-based split into train and test sets is performed, with 80% of interactions per user assigned to the train set and the remaining 20% of interactions per user assigned to the test set. The data splitting was applied via LensKit's user-based partitioning and the SampleFrac partition method, which randomly selects a fraction of test rows per user[1].

**Creating Atomic Files** The train and test sets have been transformed into the atomic file format required for use by the RecBole library[2].

**Analysis of Dataset Characteristics**

One focus of this study is to analyse the dataset characteristics and the impact Data Pruning has on the structural and distributional characteristics of the datasets. Therefore, all datasets intended for the training and testing of the Recommender algorithms were analysed under the following factors to allow further comparison following the framework employed by Deldjoo et al. [18]:

| Structural Characteristics | Distributional Characteristics |
|---|---|
| Total Number of Interactions (# Interactions) = \|Int\|<br>Total Number of Users (# Users) = \|U\|<br>Total Number of Items (# Items) = \|I\|<br>Space Size = $\|U\| \times \|I\|$ | Average Number of Interactions per User:<br>$(\text{Int}/U) = \frac{\|Int\|}{\|U\|}$<br>Average Number of Interactions per Item:<br>$(\text{Int}/I) = \frac{\|Int\|}{\|I\|}$ |

---

[1] https://lkpy.lenskit.org/0.14.4/crossfold#lenskit.crossfold.partition_users
[2] https://recbole.io/docs/user_guide/data/atomic_files.html

| | |
|---|---|
| Shape = $\frac{|U|}{|I|}$ | Gini$_{\text{Item}}$ = $1 - 2\sum_{i=1}^{|I|} \frac{|I|+1-i}{|I|+1} \times \frac{|Int_i|}{|Int|}$ |
| Density = $\frac{|Int|}{|U| \times |I|}$ | Gini$_{\text{User}}$ = $1 - 2\sum_{u=1}^{|U|} \frac{|U|+1-u}{|U|+1} \times \frac{|Int_U|}{|Int|}$ |
| Sparsity = $1 - Density$ | *These equations assume items and users are sorted according to the number of interactions associated with them (Int$_I$ and Int$_U$)* |

**Table 1 Structural and Distributional Characteristics**: *Overview of unprocessed datasets showing total counts (#), interactions (Int), users (U), and items (I), along with key characteristics (Avg. Int/User, Avg. Int/Item, shape, sparsity, Gini coefficients, and feedback type). Abbreviations are as indicated.*

Basic dataset characteristics are described by the *total number of interactions* (all user-item events), the *total number of users* (all unique individuals) and the *total number of items* (all unique entities available for interaction).

*Space Size* measures the capacity of the Interaction Matrix, allowing for the comparison of datasets in terms of the maximum number of preferences that can be collected from users.

*Shape* indicates the user-to-item ratio, which can help foresee if item-based or user-based algorithms might be advantageous.

*Density* measures the number of observed interactions in relation to the maximum number of possible interactions.

*Sparsity* describes the proportion of missing interactions in a dataset by subtracting the Density from 1.

*Average Number of Interactions per User / Item* describes the mean number of interactions a user or an item has in this dataset.

*Gini$_{Item}$* measures the rating frequency distribution for items, ranging from 0, which signifies that all items receive the same number of interactions, to 1, which indicates absolute inequality in interaction distribution, e.g., one item receiving all interactions.

*Gini$_{User}$* describes the rating frequency distribution for users, where 0 signifies that all users contribute the same number of interactions in the dataset, and 1 represents that only one user is the sole contributor of interactions.

## 4.2 Algorithm Selection

We selected a diverse set of algorithms from the LensKit [21] and RecBole [56] libraries. Random and PopScore served as baselines. Biased MF and Implicit MF represented traditional matrix-factorisation methods [48, 57], while User KNN [23] and Item KNN [19, 23] covered neighbourhood-based approaches. From RecBole, we included the BPR algorithm [40], SimpleX [33], and several deep-learning recommenders chosen with a focus on minimising energy consumption: DiffRec [50], DMF [52], and MultiVAE [32].

Hereinafter, we clustered the models into three groups.

**Group 1** consists of traditional recommender algorithms: BiasedMF, ImplicitMF, UserKNN, ItemKNN

**Group 2** is composed of modern recommender models: DiffRec, DMF, MultiVAE, SimpleX

**Group 3** holds a baseline from each library: PopScore, Random

| Library | Models | | | | | |
|---------|--------|---|---|---|---|---|
| LensKit | PopScore | Biased MF | Implicit MF | User KNN | Item KNN | |
| RecBole | Random | BPR | DiffRec | DMF | MultiVAE | SimpleX |

*Table 2: Overview of Models and Libraries included in the Experiment*

## 4.3 Training and Testing

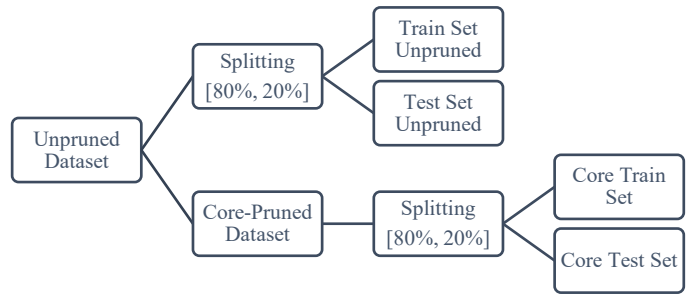

*Figure 1 Dataset Splitting for Phase 1 and 2*

In the first phase, the training and test sets were created from the respective core-pruned dataset. That means that both the training data and the test data were core-pruned.

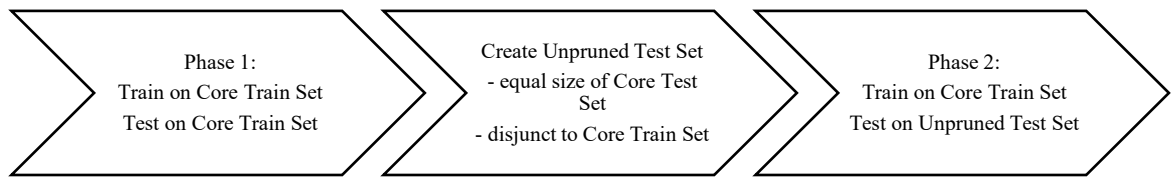

*Figure 2 Training and Testing of Recommender Algorithms employed in 2 Phases*

In the second phase, the recommenders were trained on data from the core pruned datasets, while the test data was taken from the unpruned test set. To achieve this, a test set was created from the unpruned data, which was disjoint from the core pruned training data and downsampled to the same size as the core pruned test data. By excluding possibly overlapping interactions and downsampling, a fair comparison of evaluation results on unseen data under similar conditions was enabled.

This comparison procedure was executed for each core subset of each dataset.

All recommender algorithms were trained with their default parameter as set by their respective library. The neighbourhood-based algorithms were trained with the maximum number of neighbours set to 20, and the Matrix Factorisation algorithms, BiasedMF and ImplicitMF, were trained with 50 features.

As the primary objective of this thesis was not to achieve the highest possible performance scores, but to examine changes in rank across coresets, no hyperparameter tuning was applied.

**Ranking-Based Metrics**

The evaluation of Recommender algorithms' performance on ranking tasks can be achieved through a wide variety of metrics. Research surrounding ranking tasks typically employs at least one accuracy metric, e.g. Recall or Precision, or nDCG@k in evaluation. nDCG (normalised Discounted Cumulative Gain) measures how well relevant items are ranked near the top of the recommendation list, with scores ranging from 0 for a poor list of recommendations to 1 for a perfect list of recommendations with every item at the ideal rank. Recall measures the proportion of all relevant items that appear in the list of recommendations, indicating how completely the algorithm retrieves the items of interest.

Precision reflects the proportion of recommended items that are actually relevant, assessing the accuracy of the top-ranked results. When either Precision, Recall, or nDCG is evaluated at "@k", this means the top k recommended items are considered when computing the metric.

Due to the scope of this thesis, the results will only be reported on the NDCG@10 metric. The results for Precision@10 and Recall@10 are included in the Appendix.

## 5. Results

### 5.1 Dataset Analysis

In this section, we will provide the results of analysing the dataset characteristics for all datasets. As a baseline, we will first present the characteristics of the unprocessed datasets to allow a comparison with the pre-processed, but unpruned, 0-core datasets. Subsequently, we will show the dataset characteristics for each core-pruned version of the datasets.

#### 5.1.1 Unprocessed Dataset Characteristics

| Dataset | Interactions | Users | Items | Avg. Int / User | Avg. Int / Item | Shape | Sparsity in % | Gini User | Gini Item | Feedback |
|---|---|---|---|---|---|---|---|---|---|---|
| Amazon CD | 4458901 | 434060 | 1944316 | 10.27 | 2.29 | 0.22 | 99.9995 | 0.75 | 0.50 | Explicit |
| Amazon Toys | 7998969 | 624792 | 4204994 | 12.80 | 1.90 | 0.15 | 99.9997 | 0.80 | 0.40 | Explicit |
| Gowalla | 3981334 | 107092 | 1280969 | 37.18 | 3.11 | 0.08 | 99.9971 | 0.66 | 0.54 | Implicit |
| MovieLens FL | 33832162 | 330975 | 83239 | 102.22 | 406.45 | 3.98 | 99.8772 | 0.70 | 0.95 | Explicit |
| Yelp | 5261667 | 1326101 | 174567 | 3.97 | 30.14 | 7.60 | 99.9977 | 0.63 | 0.72 | Explicit |

***Table 3 Baseline Dataset Characteristics****: Overview of the unprocessed datasets, reporting average interactions per user and per item (Avg. Int / User; Avg. Int / Item), matrix shape (users × items), sparsity, Gini coefficients for users and items, and feedback*

Analysing the unprocessed datasets confirms MovieLens as the densest. The total number of users exceeds the total number of items, and the average number of ratings, both per item and per user, are the highest of all tested datasets. While the $Gini_{User}$ value ranges in the mid ranks across datasets, the MovieLens dataset shows the highest $Gini_{Item}$ value of 0.95, exhibiting the highest inequality in rating frequency across items. This confirms the short head and long tail conformation, with a few popular items receiving large numbers of interactions and a large group of items receiving very few interactions.

The Amazon datasets are the sparsest in the experiment, exhibiting high levels of inequality in rating frequency across users. Still, they have the lowest $Gini_{Item}$ values among the tested datasets. In terms of the average number of interactions, both Amazon datasets rank in the mid-range for interactions per user but show the lowest average number of interactions per item in the experiment.

The Gowalla dataset displays the smallest Shape, with the total number of users being exceeded by the total number of items. While the average number of interactions per user is the second highest in the experiment, the average number of interactions per item is lower compared to the other datasets.

Besides MovieLens, the Yelp dataset is the only one displaying high values for both Shape and $Gini_{Item}$, as well as the second-highest average number of interactions per item. In terms of rating frequency per user, the Yelp dataset has a lower rating frequency than the compared datasets, with a $Gini_{User}$ of 0.63.

Comparing the characteristics of the unprocessed dataset to the downsampled, yet unpruned, versions of the 0-Core datasets, the relative ranking based on these statistics remained largely consistent.

## 5.1.2 Core-Pruned Dataset Characteristics

| Dataset | Interactions | Users | Items | Avg. Int. / User | Avg. Int. / Item | Space Size | Shape | Sparsity | Gini User | Gini Item |
|---|---|---|---|---|---|---|---|---|---|---|
| **0-Core Pruning** | | | | | | | | | | |
| Amazon CD | 3000000 | 376115 | 1471494 | 7.98 | 2.04 | 553450965810 | 0.26 | 99.9995% | 0.73 | 0.45 |
| Amazon Toys | 3000000 | 406436 | 1956864 | 7.38 | 1.53 | 795339976704 | 0.21 | 99.9996% | 0.74 | 0.30 |
| Gowalla | 3000000 | 104439 | 1105866 | 28.72 | 2.71 | 115495539174 | 0.09 | 99.9974% | 0.66 | 0.51 |
| MovieLens FL | 3000000 | 268342 | 34889 | 11.18 | 85.99 | 9362184038 | 7.69 | 99.9680% | 0.64 | 0.91 |
| Yelp | 3000000 | 949214 | 163998 | 3.16 | 18.29 | 155669197572 | 5.79 | 99.9981% | 0.57 | 0.73 |
| **5-Core Pruning** | | | | | | | | | | |
| Amazon CD | 2515024 | 107132 | 1308076 | 23.48 | 1.92 | 140136798032 | 0.08 | 99.9982% | 0.60 | 0.42 |
| Amazon Toys | 2499239 | 101882 | 1698279 | 24.53 | 1.47 | 173024061078 | 0.06 | 99.9986% | 0.59 | 0.28 |
| Gowalla | 2942493 | 77873 | 1089959 | 37.79 | 2.70 | 84878377207 | 0.07 | 99.9965% | 0.57 | 0.51 |
| MovieLens FL | 2718416 | 132958 | 34416 | 20.45 | 78.99 | 4575882528 | 3.86 | 99.9406% | 0.48 | 0.91 |
| Yelp | 1730318 | 129820 | 143934 | 13.33 | 12.02 | 18685511880 | 0.90 | 99.9907% | 0.45 | 0.72 |
| **10-Core Pruning** | | | | | | | | | | |
| Amazon CD | 2180293 | 55519 | 1177234 | 39.27 | 1.85 | 65358854446 | 0.05 | 99.9967% | 0.54 | 0.41 |
| Amazon Toys | 2197395 | 55611 | 1532748 | 39.51 | 1.43 | 85237649028 | 0.04 | 99.9974% | 0.53 | 0.26 |
| Gowalla | 2830783 | 61579 | 1061525 | 45.97 | 2.67 | 65367647975 | 0.06 | 99.9957% | 0.53 | 0.50 |
| MovieLens FL | 2377431 | 81663 | 34034 | 29.11 | 69.85 | 2779318542 | 2.40 | 99.9145% | 0.40 | 0.90 |
| Yelp | 1214677 | 49057 | 129533 | 24.76 | 9.38 | 6354500381 | 0.38 | 99.9809% | 0.41 | 0.71 |
| **20-Core Pruning** | | | | | | | | | | |
| Amazon CD | 1804688 | 27582 | 1019721 | 65.43 | 1.77 | 28125944622 | 0.03 | 99.9936% | 0.47 | 0.38 |
| Amazon Toys | 1830233 | 28463 | 1325512 | 64.30 | 1.38 | 37728048056 | 0.02 | 99.9951% | 0.47 | 0.24 |
| Gowalla | 2478446 | 36732 | 976549 | 67.47 | 2.54 | 35870597868 | 0.04 | 99.9931% | 0.46 | 0.49 |
| MovieLens FL | 1826386 | 41403 | 33289 | 44.11 | 54.86 | 1378264467 | 1.24 | 99.8675% | 0.32 | 0.89 |
| Yelp | 802206 | 17652 | 111692 | 45.45 | 7.18 | 1971587184 | 0.16 | 99.9593% | 0.36 | 0.68 |
| **50-Core Pruning** | | | | | | | | | | |
| Amazon CD | 1268938 | 9860 | 775156 | 128.70 | 1.64 | 7643038160 | 0.01 | 99.9834% | 0.39 | 0.34 |
| Amazon Toys | 1276790 | 10073 | 987503 | 126.75 | 1.29 | 9947117719 | 0.01 | 99.9872% | 0.39 | 0.20 |
| Gowalla | 1776980 | 14295 | 793831 | 124.31 | 2.24 | 11347814145 | 0.02 | 99.9843% | 0.40 | 0.45 |
| MovieLens FL | 894887 | 10512 | 31124 | 85.13 | 28.75 | 327175488 | 0.34 | 99.7265% | 0.24 | 0.84 |
| Yelp | 412749 | 4317 | 85747 | 95.61 | 4.81 | 370169799 | 0.05 | 99.8885% | 0.28 | 0.62 |
| **100-Core Pruning** | | | | | | | | | | |
| Amazon CD | 870960 | 4053 | 573131 | 214.89 | 1.52 | 2322899943 | 0.01 | 99.9625% | 0.33 | 0.30 |
| Amazon Toys | 855740 | 3911 | 702727 | 218.80 | 1.22 | 2748365297 | 0.01 | 99.9689% | 0.34 | 0.16 |
| Gowalla | 1156429 | 5210 | 601281 | 221.96 | 1.92 | 3132674010 | 0.01 | 99.9631% | 0.36 | 0.39 |
| MovieLens FL | 348196 | 2329 | 27076 | 149.50 | 12.86 | 63060004 | 0.09 | 99.4478% | 0.18 | 0.76 |
| Yelp | 200169 | 1182 | 61321 | 169.35 | 3.26 | 72481422 | 0.02 | 99.7238% | 0.23 | 0.54 |

*Table 4 Characteristics of the Coresets*

### 5.1.3 Structural Characteristics

The following section provides a brief overview of the observed changes in structural characteristics after pruning the datasets.

#### 5.1.3.1  Density / Sparsity

Across all pruning levels, the ranking of datasets by their density remained largely unaltered from the unpruned versions. 5-Core pruning merely switched the order of the rank-neighbours Yelp and Gowalla, similarly to 50-Core pruning for Gowalla and Amazon CD. The density of all datasets continued to increase with each Core-level.

#### 5.1.3.2  Space Size

The Space Size continuously decreased for all datasets with each stage of pruning, with only adjacently ranking datasets switching order for 10-core pruning and 50-core pruning.

#### 5.1.3.3  Shape

Across all pruning levels, the Shape values of all datasets decreased continuously, with the largest difference between the unpruned and 5-core pruned datasets reported.

#### 5.1.3.4  Average Number of Interactions per Item

The average number of interactions per item varied for the unprocessed datasets, with values ranging from 1.53 for Amazon Toys up to 85.99 for MovieLens. The span between the maximum and minimum values stayed considerably large across all pruning levels, with the rank of datasets by this characteristic entirely unchanged.

#### 5.1.3.5  Average Number of Interactions per User

The ranking by Average Number of Interactions per User fluctuated notably more than per Item. Pruning at the lower level of 5-core, the sparse Amazon Toys dataset, ranked fourth for 0-core, switched ranks with the formerly ranked second MovieLens dataset when ranked in descending order by Average Number of Interactions per User.

20-Core pruning resulted in every dataset switching rank with an adjacent dataset, except for the Gowalla dataset, which consistently displayed the highest average number of interactions per user across all core levels. Pruning at thresholds above 50 relegated the MovieLens dataset, which ranked second highest in the unpruned version, to the last rank of all datasets.

### 5.1.4 Distributional Characteristics

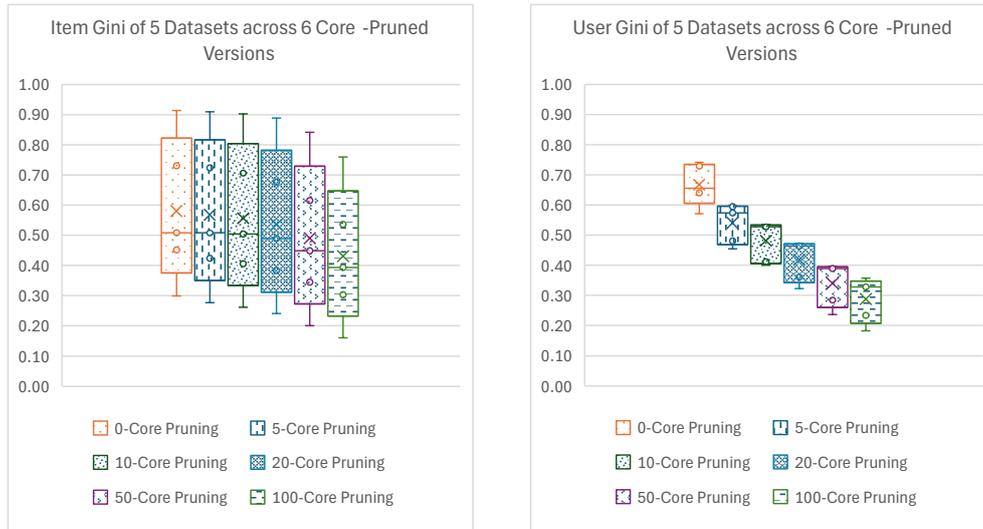

***Figure 3:*** *Boxplots showing the distribution of **Gini Coefficients** for users and items across pruning levels, where 1 signifies total inequality in interaction distribution and 0 stands for a uniform distribution of interactions*

*5.1.4.1        $Gini_{User}$*

All datasets had a $Gini_{User}$ value of more than 0.5 for the downsampled but unpruned version, with the Amazon datasets displaying the maximum values in the experiment of 0.7. For all core-pruned versions, the highest values continued to be those of the Amazon datasets, and MovieLens had the lowest values for all thresholds above 5. The Amazon and Gowalla datasets showed similar decreasing trends and values across all pruning levels, as did Yelp and MovieLens, but their $Gini_{User}$ values were lower throughout all levels.

*5.1.4.2        $Gini_{Item}$*

The values for $Gini_{Item}$ consistently stayed in the same order for all pruning levels, with MovieLens displaying the maximum value of 0.9 for the unpruned version and 0.8 for the 100-core pruned version. While each was about 0.2 lower than MovieLens, both the Yelp and Gowalla datasets showed a similar trend, reaching minimum values of 0.5 and 0.4, respectively, in the 100-core version for $Gini_{Item}$. Overall, the datasets displayed a drop of a maximum of 0.2 from 0-core to 100-core pruned in terms of $Gini_{Item}$, thus a much lower fluctuation than $Gini_{User}$.

**Relative Retention Rates**

In this section, we provide an in-depth analysis of the impact of data pruning with increasing core thresholds on dataset characteristics, with a focus on the extent to which the users, items, and interactions of the unpruned dataset are retained in the pruned datasets.

*5.1.4.3    User Retention Rate*

Across all five examined datasets, pruning has affected the relative retention of users as early as the first pruning level of 5-core. Only one dataset, namely Gowalla, retained more than half of its users compared to the baseline, with a 75% relative retention rate. The lowest user retention rate for 5-core pruning was observed for the Yelp dataset, retaining only 14% of its user base after the first level of pruning was applied. Both Amazon datasets, CD & Vinyl and Toys & Games, reported a relative user retention under 30%, with only a 3% difference between them. The MovieLens dataset retained 50% of the users compared to the unpruned version.

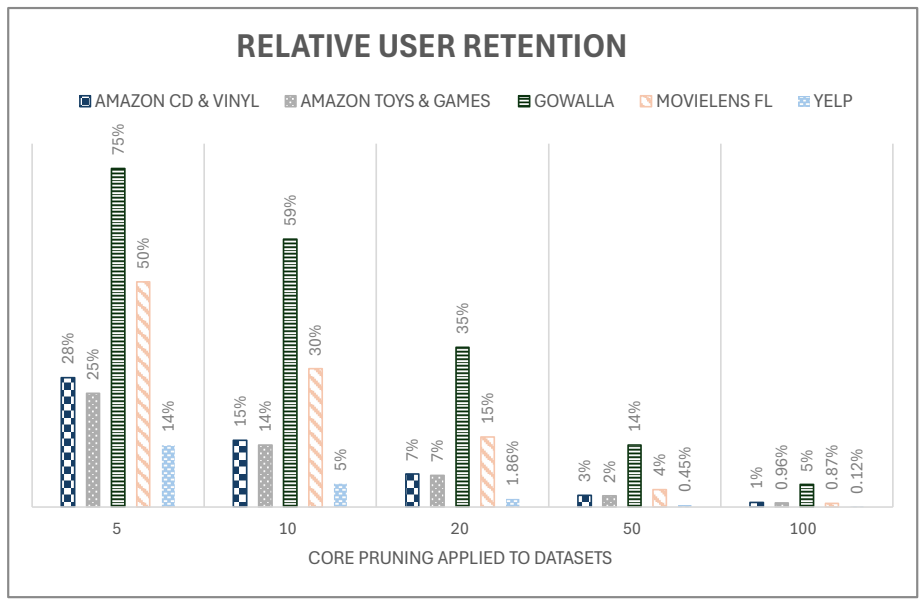

*Figure 4 User Retention Across Core Levels: Grouped bar chart showing, for each pruning level, the proportion of users retained in each dataset relative to its 0-core (unpruned) set.*

Applying 10-core pruning to the datasets only left the Gowalla dataset with more than half of the users, while MovieLens dropped to less than a third of the original user-base, with 30% of users retained. The remaining datasets reported retention rates below 15%, with Yelp dropping as low as 5% relative user retention.

The commonly executed 20-core pruning reduced the user numbers for the MovieLens dataset by half compared to the 10-core version, with a user retention rate of 15%. Both Amazon datasets dropped to values below 10% and the Yelp dataset retained merely 1.86% of its original user base. Gowalla continued to maintain the largest relative share of users throughout the remaining core levels of 20, 50, and 100, with 35%, 14%, and 5% respectively.

With the exception of the Gowalla dataset, pruning at the core level of 50 dropped all remaining datasets to below 5% and pruning at the highest examined level of 100 dropped to below 1% of users retained compared to the unpruned dataset. The lowest user retention rate was examined for the 100-core pruned Yelp dataset, reporting 0.12% of the original user base in the coreset.

*5.1.4.4       Item Retention Rate*

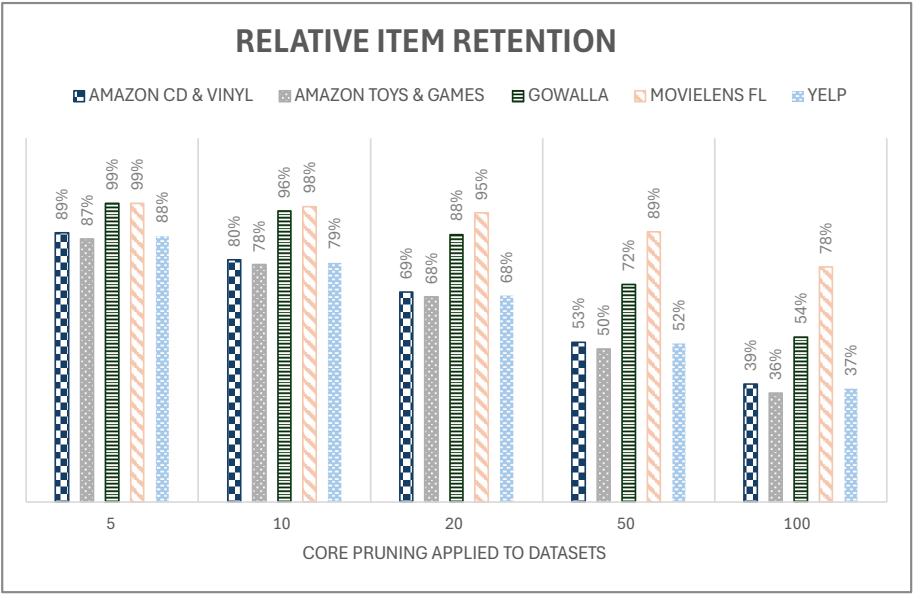

*Figure 5 Item Retention Across Core Levels*: *Grouped bar chart showing, for each pruning level, the proportion of items retained in each dataset relative to its 0-core (unpruned) set.*

Pruning datasets affects the relative retention of items compared to the unpruned datasets. Similar to the user retention rate, the item retention rate dropped continuously as the pruning levels rose, but for items, the retention rates stayed higher than for users.

Examining the 5-core pruned datasets, all datasets retained more than 85% of items compared to the unpruned dataset, with MovieLens and Gowalla only dropping by 1% to 99% of item retention. Both Amazon datasets and the Yelp dataset retained around 80% of items.

10-core pruning the datasets resulted in item retention rates dropping by between 1% for the MovieLens dataset and 9% for both Amazon datasets, and the Yelp dataset.

The popular 20-core pruned versions of the datasets continued to show similar retention rates for the two Amazon datasets and the Yelp dataset, dropping to just below 70% item retention. The MovieLens dataset retained the most items with 95% item retention, and the Gowalla dataset retained 88% of items.

Pruning with a core level of 50 left the Amazon datasets and the Yelp dataset with about 50% of the original items. The Gowalla dataset reported a retention rate of 72% and the MovieLens dataset dropped just below 90% of item retention.

The trend continued for Core 100: the Yelp and Amazon datasets dropped to rates between 36% and 39%, while Gowalla retained about half of the items compared to the unpruned dataset. The MovieLens dataset showed an item retention rate of 78% for the highest level of pruning examined in the study.

*5.1.4.5   Interaction Retention Rate*

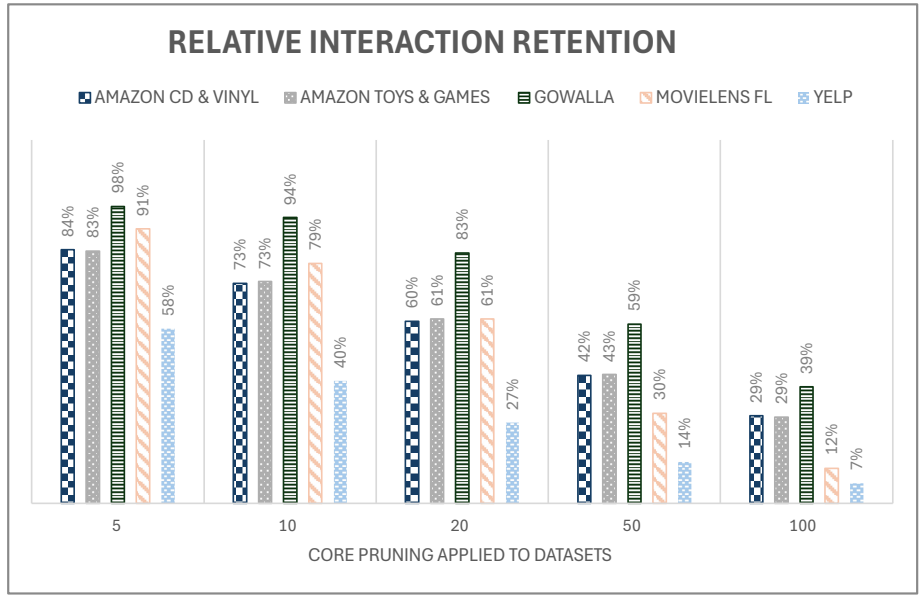

*Figure 6 Interaction Retention Across Core Levels: Grouped bar chart showing, for each pruning level, the proportion of interactions retained in each dataset relative to its 0-core (unpruned) set.*

Analysis of the retained interactions after pruning the datasets revealed differences in the extent to which the datasets were affected by the pruning levels. For the first level of pruning to core 5, the majority of datasets remained with more than 80% of interactions, with Gowalla showing the highest interaction retention rate of 98%. The Yelp dataset was reduced by 42% through 5-core pruning, retaining 58% of the interactions.

This trend continued for 10-core pruning: both Amazon datasets and the MovieLens dataset retained around 70-80% of their interactions, while Gowalla retained the biggest relative portion of 94% of interactions. Pruning the Yelp dataset with a threshold of 10 dropped the interaction retention below the halfway mark, with 40% of interactions retained in the 10-coreset.

Pruning with 20-core levels, all datasets dropped by between 10-20% of interaction retention rates, with Yelp reaching as low as 27% and Gowalla retaining 83% of interactions.

The 50-core pruning had the most significant impact on the MovieLens dataset, resulting in a 30% drop in interaction retention, which led to lower retention rates compared to the Amazon datasets, at 42% and 43%.

Pruning with a threshold set to 100 only retained more than a third of the original interactions in the Gowalla dataset and displayed retention rates below 15% for both the MovieLens and Yelp datasets.

## 5.2 Algorithm Performance Analysis

This chapter is organised into three parts to provide a comprehensive overview of the results obtained by training and testing 11 recommender algorithms on progressively pruned versions of five datasets. The first section presents the overall performance trends observed for the three algorithm groups of traditional, modern, and baseline algorithms. The second section reports the results for each group when both training and testing were conducted on the pruned datasets. The final section presents the outcomes of training performed on pruned datasets while testing on unpruned datasets.

### 5.2.1 Overall Performance Trends

This section summarises the overall performance patterns observed for three algorithm groups across all pruning levels, when trained and tested on pruned data. Group 1 consists of traditional algorithms: User KNN, Item KNN, Biased MF, Implicit MF and BPR. Group 2 comprises the modern algorithms we have selected: DiffRec, DMF, MultiVAE, and SimpleX. Group 3 comprises the two baseline algorithms chosen for the experiments: Random and PopScore.

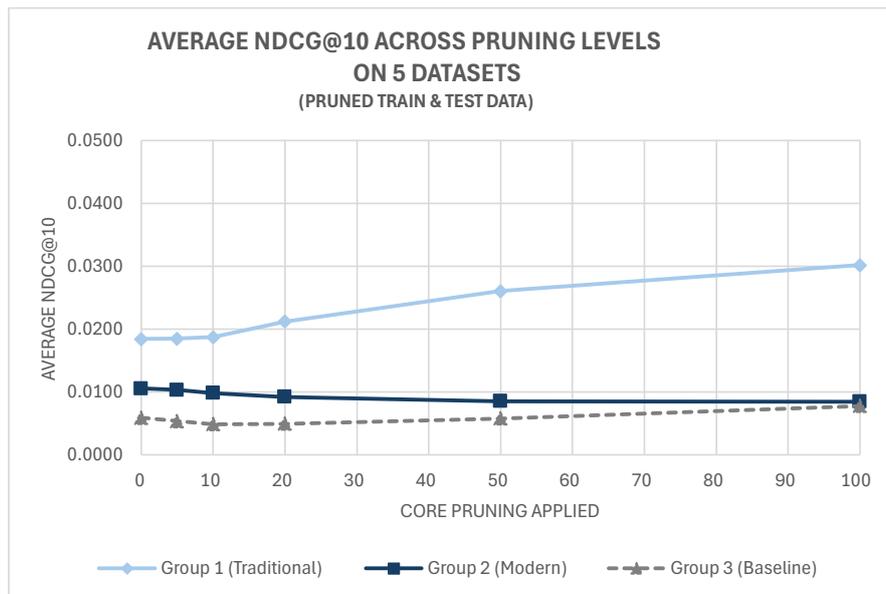

*Figure 7 Average nDCG@10 Across Pruning Levels: Average nDCG@10 of traditional, modern, and baseline algorithms on five datasets, trained and tested on equally pruned data.*

For the traditional algorithms (Group 1), the average nDCG@10 measured across five datasets showed consistent values for 5-core and 10-core pruning, with an overall increase in nDCG@10 for the 50-core and 100-core pruned versions of the datasets. The second group, modern algorithms, on average, did not improve performance with progressively pruning the datasets. The baseline algorithms in Group 3 overall showed almost no change in nDCG@10.

While the averaged nDCG@10 for the three groups displayed a maximum difference of 0.01, the performance of the algorithms in these groups showed notable fluctuations.

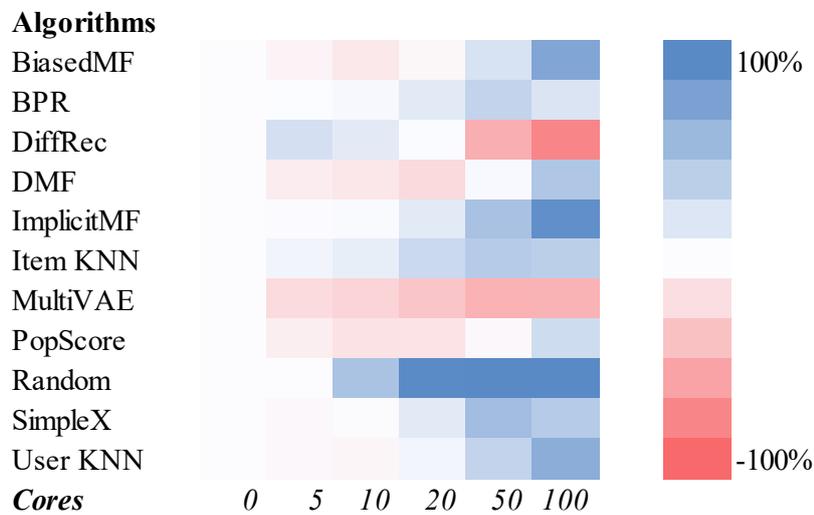

*Figure 8 Average Relative Change (%) in nDCG@10 vs. 0-core Baseline*: Heatmap showing the percent change in nDCG@10 at each pruning level relative to the unpruned (0-core) performance; positive values (blue) indicate improvement, negative values (red) indicate decline.

Comparing the averaged nDCG@10 values per algorithm showed both positive and negative relative changes compared to training and testing on the 0-core version of the datasets. Implicit MF (+ 96%) and Random (+ > 100%) displayed high relative differences to their results on the 0-coresets, while DiffRec had an average relative change of -80% for 100-core pruned datasets. This variance in Average Relative Change highlights pronounced differences between the reactions to pruning levels, prompting a more in-depth analysis of algorithm performance. In the following subsection, we will analyse the performance of the algorithms within the groups in more detail.

### 5.2.2 Training and Testing on Pruned Datasets

In this section, we will provide an overview of the algorithms' performance, both trained and tested on the datasets in the core-pruned versions (Phase 1). The algorithms are grouped into three categories: Group 1, which includes traditional recommender algorithms such as User KNN; Group 2, comprising modern algorithms, e.g., DiffRec and DMF; and Group 3, consisting of the two included baselines: Random and PopScore.

#### 5.2.2.1 Group 1: Traditional Recommender Algorithms Trained and Tested on Pruned Data

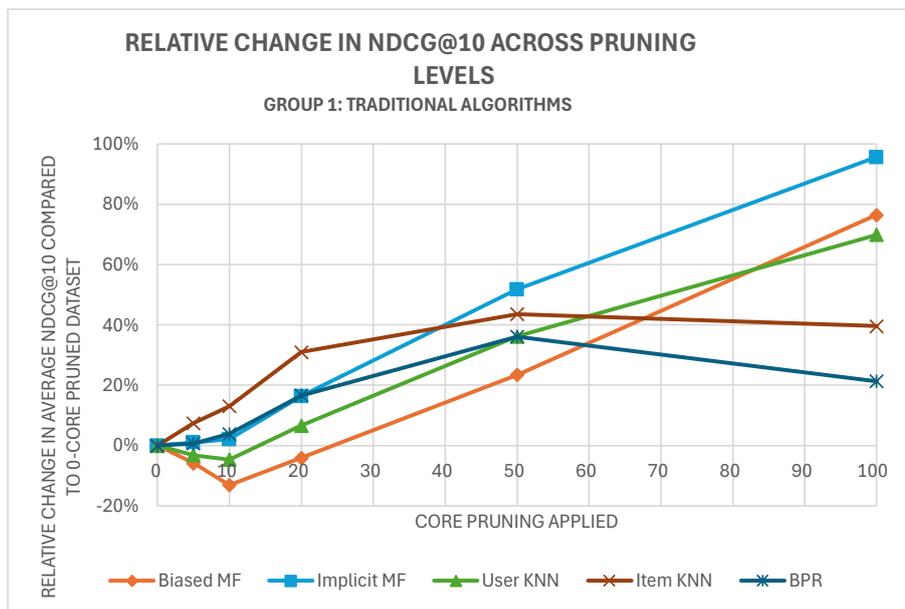

*Figure 9 Relative Change in nDCG@10 Across Pruning Levels for Traditional Algorithms:* Percentage change in average nDCG@10 of BiasedMF, ImplicitMF, UserKNN, ItemKNN, and BPR compared to their performance on the unpruned (0-core) datasets.

Measuring the relative change in average nDCG@10 between each core version and the baseline of 0-core, deviations from the baseline were reported for all included algorithms (**Figure 9**).

Biased MF dropped by 6% for 5-core and 13% for 10-core pruning, when compared to the average nDCG@10 achieved on the 0-core datasets. For 50-core and 100-core pruning, Biased MF increased in average nDCG@10 by 23% and 77% respectively.

Implicit MF displayed negligible changes in performance for 5- and 10-core pruning (< 3%), but improved performance for the remaining pruning levels of 20, 50 and 100. The highest measured values in relative change of this group were calculated for Implicit MF at 50-core (52%) and 100-core (96%).

Across the five datasets, User KNN experienced an average decrease in nDCG@10 for 5-core pruning (-3%) and 10-core pruning (-5%), while showing notable improvements in performance for the 50-core (36%) and 100-core (70%) pruned datasets.

Item KNN displayed the highest increase for the first two pruning levels of 5 (7%) and 10 (13%), ranked second best for 50-core pruning (44%), but showed less improvement for 100-core pruning than the majority of algorithms in the group (40%).

The final algorithm of Group 1, BPR, showed negligible improvement for the first two levels of pruning, with under 5% improvement for both 5-core and 10-core pruning. The relative change increased for 20-core pruning (17%), 50-core pruning (36%), and 100-core pruning (21%).

### 5.2.2.2 Group 2: Modern Recommender Algorithms Trained and Tested on Pruned Data

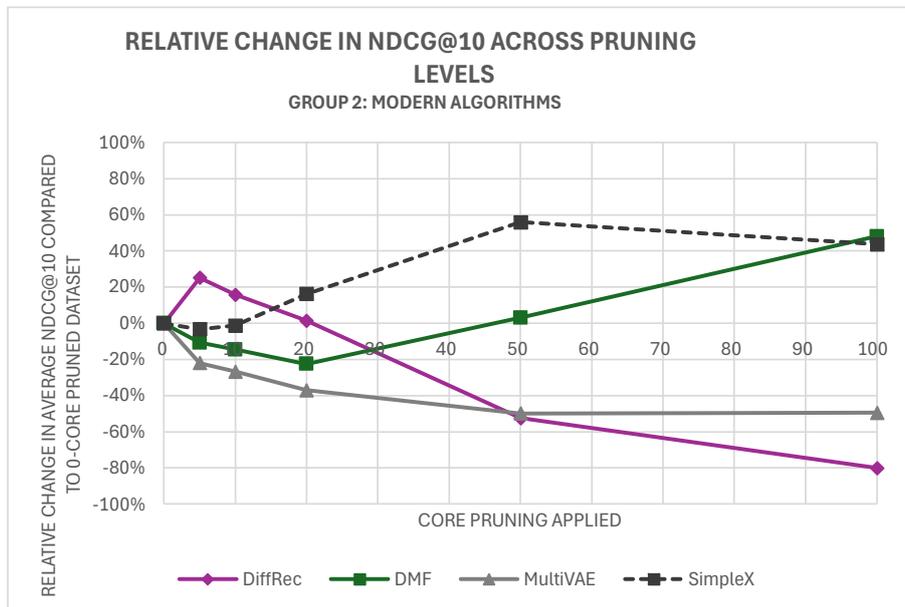

*Figure 10 Relative Change in nDCG@10 Across Pruning Levels for Modern Algorithms:* Percentage change in average nDCG@10 of DiffRec, DMF, MultiVAE and SimpleX compared to their performance on the unpruned (0-core) datasets.

The second group of algorithms, comprised of the modern recommenders DiffRec, DMF, MultiVAE, and SimpleX, deviated from the traditional recommender group in multiple pruning stages.

Comparing the average nDCG@10 across five datasets, DiffRec improved by 25% relative to the averaged performance on the 0-core versions of the datasets for 5-core pruning. The second pruning level of 10-core yielded a 16% increase, while 20-core pruning displayed negligible changes in performance (1%). For both 50-core and 100-core pruning, DiffRec's performance decreased by 52% and 80%, respectively.

In contrast to DiffRec, DMF displayed a decline in performance for pruning cores 5, 10 and 20, a negligible increase for 50-core pruning (< 5%) and overall improved most on the 100-core pruned dataset by 48%.

MultiVAE was the only observed algorithm in this group to exhibit a continuous decline in average nDCG@10 across all datasets for each pruning level. Even 5-core and 10-core pruning decreased performance compared to the unpruned dataset by over 20%, and pruning at levels equal to or higher than 50 reduced the performance by 50% on average.

As the final algorithm of this group, SimpleX displayed minor relative changes compared to 0-core datasets for 5-core and 10-core pruning (< -5%), but consistently increased in average nDCG@10 for 20-core (16%), 50-core (56%) and 100-core (44%) pruned datasets. The relative change for 20-core and 50-core was the highest relative improvement observed in this group.

### 5.2.2.3 Group 3: Baseline Recommender Algorithms Trained and Tested on Pruned Data

Group 3, containing only the two baseline algorithms PopScore and Random, displayed the overall lowest nDCG@10.

Random only produced a measurable value for nDCG@10 on the MovieLens dataset for the 0-core and 5-core versions, as well as the Yelp dataset for the 10-core versions. Each subsequent pruning stage added another dataset to the set of those generating a measurable value for nDCG@10, resulting in relative improvements of 350% for 50-core and 850% for 100-core. Executing the Random algorithm on the Amazon Toys & Games dataset yielded no observable value for nDCG@10 regardless of pruning intensity.

In contrast to Random, training and testing the PopScore algorithm achieved measurable nDCG@10 values above zero across all core-pruned versions of all datasets. While improvements were observed for the 5-core version of Gowalla, the overall nDCG@10 declined by 9% for the 5-core pruned versions of the datasets. Pruning at 10 and 20 decreased the average performance by 18% and 17%, respectively, compared to the 0-core pruned datasets. Only 100-core pruning achieved an improved average nDCG@10 (29%).

### 5.2.3 Comparison of Testing on Pruned and Unpruned Testsets

Building on the previous analyses, this section focuses on the setup for comparison. For clarity, we will distinguish between two experimental phases: Phase 1, also represented by the previous section of the Results chapter, refers to training and testing the algorithms on pruned datasets. Phase 2 denotes the setting in which the training was executed on the pruned training sets, whereas the testing was performed on the unpruned test sets. The 0-coresets were excluded from Phase 2, as the training data in these coresets was not pruned.

To facilitate a direct comparison of the influence data pruning has on performance scores under these two evaluation setups, a ranking of the algorithms by descending performance (measured in nDCG@10) was created and visualised in a heatmap (**Figure 11**). Rank 1 (dark blue) signifies the best performance, as measured by the average nDCG@10 of all algorithms on this particular coreset of the respective dataset. In contrast, Rank 11 (dark red) marks the lowest measured performance score.

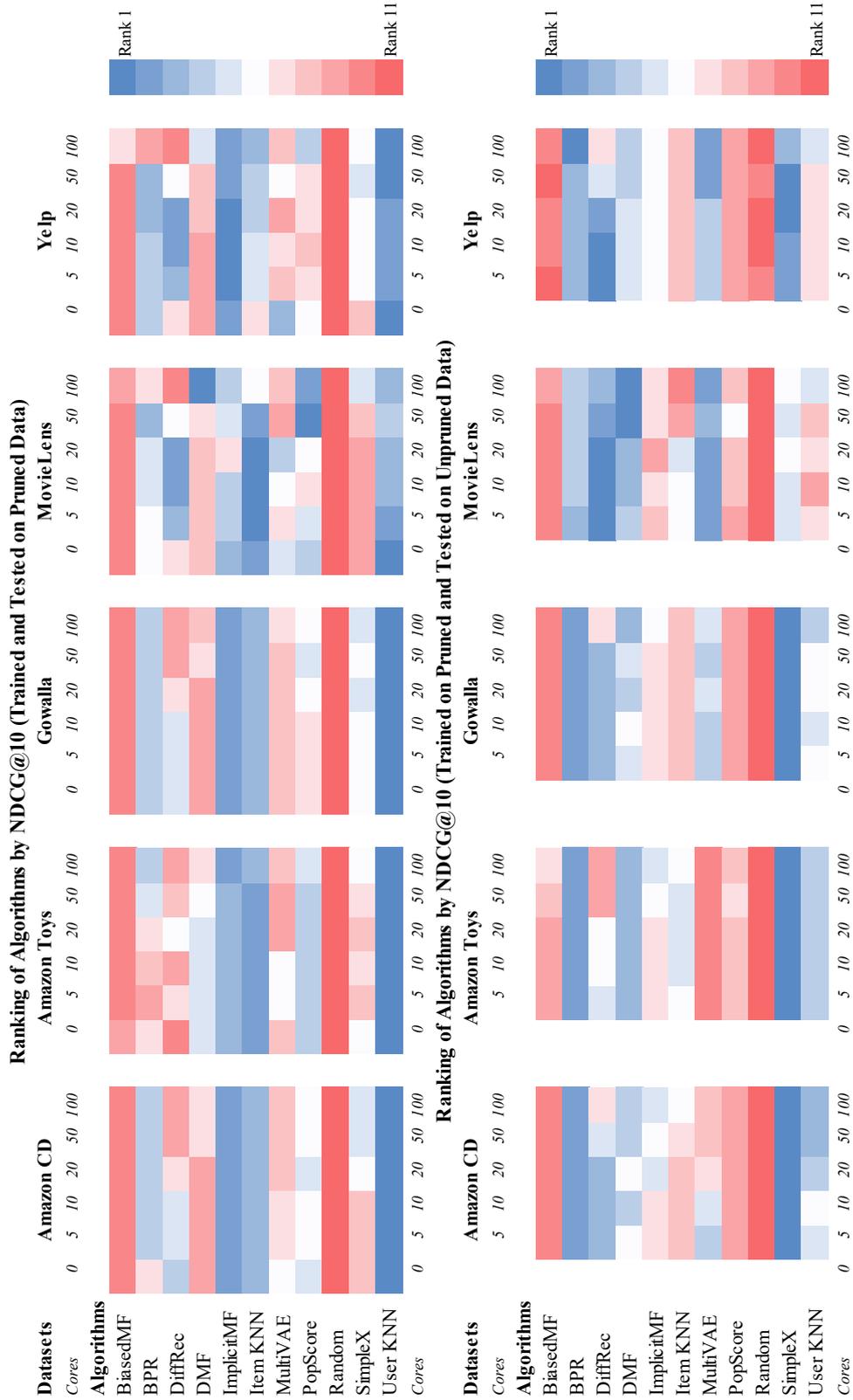

*Figure 11 Algorithm Ranking Across Two Evaluation Phases:* Two-part heatmap showing algorithm ranks (based on nDCG@10) across five datasets and six pruning levels. The upper panel depicts Phase 1 (training and testing on equally pruned data), while the lower panel shows Phase 2 (training on pruned data and testing on unpruned data).

In Phase 1, User KNN ranked highest for the entirety of three datasets, on several coresets of the remaining two datasets (MovieLens, Yelp) and displayed the highest consistency in ranking of both traditional and modern algorithms. On the Yelp dataset, User KNN was outperformed by Implicit MF on three coresets: 5-core, 10-core and 20-core. The MovieLens dataset exhibited more fluctuations for the higher ranks, with User KNN ranking third or fourth on all coresets above 5-core, and Item KNN rising to first rank for 5-core, 10-core, and 20-core pruning. DiffRec ranked second for the 5-core and 10-core subset, and DMF took the lead for the 100-core pruned subset of MovieLens. PopScore ranked highest on the 50-core subset of MovieLens.

Phase 1 was characterised by consistent top rankings of User KNN, Item KNN, and Implicit MF, with Item KNN and Implicit MF exhibiting rank fluctuations primarily across entire datasets and to a lesser extent on singular coresets.

Algorithms at mid-level ranks displayed a higher frequency of rank changes, often within a dataset. One example of this behaviour would be MultiVAE across the MovieLens dataset, where its rank fluctuated between fourth and ninth place. DiffRec achieved varying ranks both across entire datasets and coresets, with ranking position changing from second to as low as tenth across the Yelp dataset and almost a different rank for every coreset on MovieLens (only 10-core and 20-core ranked second).

The only algorithm without any fluctuations in ranking in Phase 1 was Random, which continuously ranked last ($11^{th}$ out of 11 algorithms). Almost no variation in rank was also observed for Biased MF, which ranked $10^{th}$ on all but three coresets in Phase 1.

In summary, rank changes were observed frequently in Phase 1, both per dataset (e.g., SimpleX on MovieLens) and per coreset, as evidenced by PopScore for both MovieLens and Yelp.

In comparison to Phase 1, the top contender User KNN displayed both more frequent and more extreme rank changes in Phase 2: While the lowest rank in Phase 1 was a singular fourth rank on 50-core pruned MovieLens, User KNN held a rank equal to or below sixth for 48% of coresets in Phase 2 (12 of 25). As the second neighbourhood-based algorithm, Item KNN also consistently showed a lower rank than in Phase 1, ranking overall eighth for both entire datasets and achieving its best ranking of fifth.

Implicit MF, which even ranked first for three coresets in Phase 1, mostly ranked fifth to seventh, and on the 20-core subset of MovieLens, even ranked ninth. Biased MF was ranked $10^{th}$ for 68% (17 of 25) of coresets in Phase 2, with a slightly higher frequency of rank changes on Amazon Toys & Games and the Yelp dataset.

BPR was observed to rank fourth for 68% (17 of 25) in Phase 1, compared to ranking second for 60% of coresets in Phase 2. On the 100-core pruned subset of Yelp, BPR even ranked best of all algorithms.

MultiVAE exhibited difficulties in recommending on the Amazon Toys dataset in Phase 2, consistently ranking $10^{th}$ place. While the performance was in the lower ranges for the Amazon datasets, on the remaining three datasets, MultiVAE consistently outperformed the majority of algorithms: On Yelp, MultiVAE held the second rank for 50-core and 100-core, on MovieLens for all but 50-core, and on Gowalla, the rank varied between fourth and fifth.

DiffRec placed higher in the ranks in Phase 2 compared to the first test phase, still overall performing best on Yelp and MovieLens, as it did in Phase 1. Rank changes were observed less frequently compared to Phase 1. DiffRec performance notably declined with each progressive pruning stage on the Amazon Toys & Games dataset, ranking fifth on 5-core, sixth on 10-core and 20-core, and ninth on 50-core and 100-core. However, the absolute nDCG@10 values were observed to be higher in Phase 2 than in Phase 1. On the 5-core pruned dataset, DiffRec achieved a nDCG@10 of 0.0153 in Phase 1, compared to 0.1155 in Phase 2.

DMF was the only algorithm showing improvements in rank by performance on the more severely pruned version of MovieLens. Contrary to the behaviour examined for the remaining algorithms, DMF achieved a higher rank for each core level on MovieLens, with a first rank for both the 50-core and 100-core models. On Amazon Toys & Games, no rank variation was observed for DMF, which consistently ranked third best for all coresets.

SimpleX achieved an undisputed first rank on three datasets (both Amazon and Gowalla), ranked between fifth and sixth on all coresets of MovieLens and showed minor fluctuations within the top 3 ranks on the Yelp coresets. Its average nDCG@10 across all coresets was 1276% higher than in Phase 1, and the absolute nDCG@10 values were the highest observed in the entire experiment.

| Average nDCG@10 across 5 Pruning Levels (Trainset Pruned, Testset Unpruned) | | | | | |
|---|---|---|---|---|---|
| **Cores** | **5** | **10** | **20** | **50** | **100** |
| Biased MF | 0.0001 | 0.0001 | 0.0002 | 0.0001 | 0.0001 |
| Implicit MF | 0.0013 | 0.0013 | 0.0020 | 0.0014 | 0.0014 |
| User KNN | 0.0018 | 0.0018 | 0.0025 | 0.0019 | 0.0017 |
| Item KNN | 0.0010 | 0.0011 | 0.0018 | 0.0010 | 0.0009 |
| BPR | 0.0628 | 0.0509 | 0.0364 | 0.0213 | 0.0136 |
| DiffRec | 0.1155 | 0.0705 | 0.0194 | 0.0043 | 0.0022 |
| DMF | 0.0078 | 0.0073 | 0.0056 | 0.0059 | 0.0048 |
| MultiVAE | 0.0102 | 0.0087 | 0.0087 | 0.0045 | 0.0041 |
| SimpleX | <u>0.2216</u> | <u>0.1702</u> | <u>0.1123</u> | 0.0540 | 0.0253 |
| PopScore | 0.0006 | 0.0005 | 0.0011 | 0.0005 | 0.0003 |
| Random | 0.0001 | 0.0000 | 0.0000 | 0.0001 | 0.0000 |

*Table 5 Average nDCG@10 of Algorithms over 5 Datasets Across 5-Core Pruning Levels in Phase 2,* with the highest overall values in the experiment underlined

In summary, the traditional algorithms in Group 1 overall displayed a lower performance in Phase 2 compared to Phase 1, while the modern algorithms in Group 2 partially exhibited substantial increases in measured nDCG@10 values. Notably, SimpleX and DiffRec showed an improved performance under the second test configuration when tested on unpruned test sets. In comparison with the 5-core baseline in Phase 2, each additional pruning level resulted in a notable performance decrease observed for all Group 2 (Modern) algorithms. The Interpretation and Discussion chapter will elaborate on this particular finding.

# 6. Discussion and Interpretation

## 6.1.1 Dataset Characteristics

In this section, we examine how progressive pruning influenced key characteristics of the datasets. We begin by analysing the impact of data pruning on structural properties, e.g. density and the average number of interactions per user and per item, as well as distributional characteristics, such as the Gini coefficients of users and items. Secondly, we focus on retention rates, discussing the extent of data loss observed across increasing levels of pruning.

### 6.1.1.1 Structural Dataset Characteristics

Our analysis of the datasets' characteristics revealed that pruning affected all of the examined datasets, though to varying extents. The removal of users and their interactions from a dataset resulted in an unsurprising decrease in the total number of interactions, users, and items for each core-level pruning applied to the datasets. Subsequently, structural characteristics dependent on at least one of these three total numbers were equally impacted by pruning: Space Size, which describes the maximum possible interactions calculated as users × items, and Shape, denoting the relation of the total number of users to the total number of items, continuously decreased with each pruning stage.

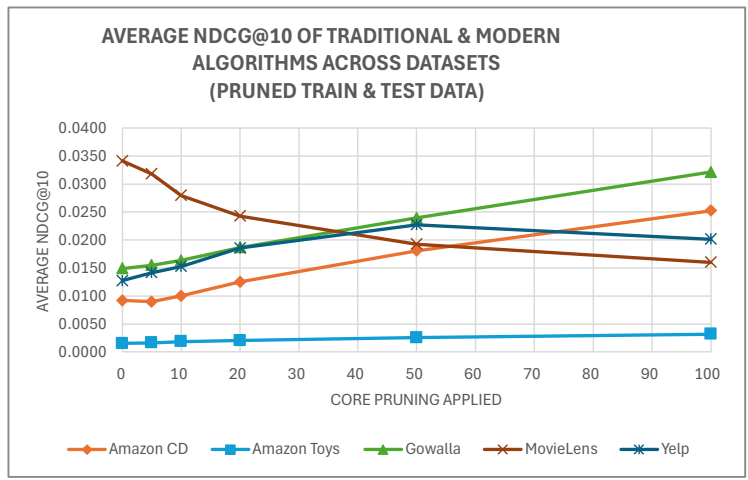

*Figure 12 Average nDCG@10 of Traditional and Modern Algorithms Across Datasets:* Average nDCG@10 for five datasets at different core-pruning levels when training and testing are both performed on equally pruned data.

Density, a key factor in explaining performance variance [1], was observed to increase with each core level across all examined datasets in our study. While the sparser datasets, Amazon CD and Amazon Toys, showed improved performance for both the traditional and baseline algorithms on coresets with higher density, this was not the case for MovieLens. Being the densest dataset in the experiment, MovieLens exhibited distinct behaviour: despite increasing density with progressive pruning, both groups of traditional and modern algorithms showed no performance gains. On the contrary, for most core levels, declining performance was observed on the MovieLens datasets as seen in **Figure 12**. These findings indicate that density alone cannot account for the performance variance observed under progressive pruning.

Researchers have recently identified "Interactions per User" as another key dataset characteristic when examining performance variance [18]. In our experiment, the Yelp dataset showed the most significant rise in Average Interactions per User, increasing by 322% (from 3.16 to 13.33) for 5-core pruning up to 5258% (from 3.16 to 169.35) for 100-core pruning. In Phase 1, the average nDCG@10 indeed increased for the majority of algorithms alongside the Average Interactions per User on the Yelp dataset, with a noticeable drop in performance for 100-core pruning. A similar trend was observed on the Gowalla dataset, which showed the overall smallest increase in Interactions per User – from 32% (from 28.72 to 37.79) at 5-core to 673% (from 28.72 to 221.96) at 100-core pruning. Algorithm performance did not decline even for the 100-core subset of Gowalla, suggesting that additional factors influence both the dataset characteristics and algorithm performance.

The impact of pruning on Average Interactions per Item was less pronounced than on interactions per user, but differed considerably among datasets. At 5-core pruning, Gowalla remained unaffected, while Yelp displayed a 34% (from 18.29 to 12.02) reduction. Across all pruning levels, MovieLens and Yelp experienced the most potent effects, with decreases of up to 85% (from 85.99 to 12.86) and 82% (from 18.29 to 3.26), respectively, for 100-core pruning. This division between moderately and heavily affected datasets aligns with changes in Space Size: MovieLens and Yelp consistently exhibited the lowest values, whereas Gowalla and both Amazon datasets retained much higher portions of Space Size across pruning levels.

Analysis of $Gini_{Item}$ divided the datasets into the same two groups. Amazon CD, Amazon Toys, and Gowalla consistently exhibited lower $Gini_{Item}$ values, indicating a pattern closer to equal interaction distribution across items. In contrast, MovieLens and Yelp displayed the highest and second-highest $Gini_{Item}$ values at every pruning level, denoting a highly uneven distribution of interactions. MovieLens, in particular, is known for a short-head, long-tail distribution, where a few popular items attract many interactions, while most items receive little interaction [8, 18, 27]. This trait is reflected in its $Gini_{Item}$ scores. Overall, $Gini_{Item}$ scores varied little across pruning stages and preserved the dataset ranking observed for the 0-coresets through to the 100-coresets.

$Gini_{User}$ exhibited trends similar to $Gini_{Item}$, with scores decreasing at each pruning level. However, the difference between unpruned datasets and the 100-coresets was substantially larger than the corresponding $Gini_{Item}$ differences. The most significant decline was observed for MovieLens, where the 100-coreset showed a drop of -0.46 compared to the 0-coreset. All other datasets followed the same pattern, though less markedly, with $Gini_{User}$ reductions ranging from -0.24 to -0.40. This outcome aligns with expectations: progressively excluding low-activity users leaves a population of generally high-activity users, thereby driving the interaction distribution closer to uniformity.

In summary, pruning affected all structural characteristics of the datasets, though not in uniform severity. Density increased for all datasets; however, its link to performance was inconsistent, as demonstrated by the declining performance scores on the already dense MovieLens dataset. Interactions per User overall increased for each dataset per core level, yet yielded varying outcomes as displayed in the comparison of Yelp and Gowalla: Gowalla, with the lowest relative increase in interactions per user, showed a steady rise in algorithm performance, while Yelp (highest relative increase in interactions per user) declined after 50-core pruning. Additional properties, such as the Gini measures for items and users, highlighted the impact on user-centred characteristics, while showing that item characteristics are affected to a lesser extent.

*6.1.1.2    Distributional Characteristics: Retention of Interactions, Items and Users*
Pruning affected the datasets' structural properties to differing degrees, while similar trends of varying intensity were frequently observed. A comparable pattern emerged examining the retention rates of interactions, items, and users across pruning levels. In this context, retention is defined as the proportion (in per cent) of the original dataset that is preserved in a coreset. User retention, therefore, describes the percentage of users retained in a coreset compared to the 0-core baseline, and this definition applies equally to item and interaction retention.

Among the three retention measures, user retention showed the most pronounced variation, differing substantially across both entire datasets and core levels. In our analysis, we observed that the impact of data pruning on datasets varies already at low core levels: 5-core pruning revealed a gap of 61% between the datasets with the highest and lowest user retention rates.

Sun et al. [47] reported 5 and 10 as common core-levels observed for 60% of experiments applying data pruning. Comparing the Gowalla and the Yelp datasets side by side revealed how varying the impact of each core level can be on two different datasets (**Figure 13**). Pruning Yelp at 5-core removed 86% of users, while retaining 58% of interactions. Meanwhile, Gowalla marked a 25% reduction in users, losing merely 2% of total interactions in the dataset. 5-core pruning removed 86% of Yelp's users while preserving 58% of its interactions, whereas Gowalla experienced a 25% loss in users, retaining 98% of its interactions. This difference in effect stems from structural differences between the two datasets: Yelp contains a significantly higher number of users than items, with the lowest $Gini_{User}$ among the unpruned datasets. In contrast, Gowalla's $Gini_{User}$ is insignificantly higher, but it contains more items than users. Therefore, pruning low-activity users removes a smaller share of the total interactions for Gowalla, causing a steadier decline in user retention across pruning levels.

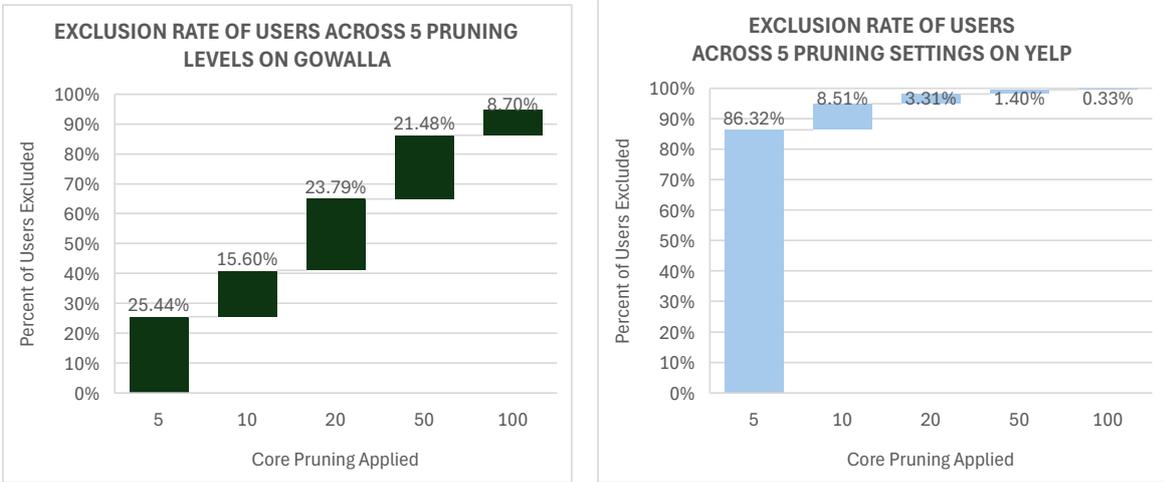

*Figure 13 User Exclusion Rates Across Pruning Levels for Gowalla and Yelp:* *Waterfall charts showing the percentage of users removed at each core-pruning level (5, 10, 20, 50, 100) for the Gowalla (left) and Yelp (right) datasets, illustrating their contrasting user-retention patterns per level.*

The use of 20-core pruned datasets (e.g., MovieLens 100k, MovieLens 1M, …) is a common design choice in recommender research, but Beel and Brunel [8] already established in 2018 that this results in the exclusion of 42% of users compared to an unpruned version of the dataset. In our experiments, this rate was considerably higher: the examined datasets experienced user loss ranging from 65% for Gowalla, over 93% (both Amazon datasets), up to 98.14% (Yelp) when 20-core pruning was applied. Even the popular MovieLens dataset retained merely 15% of users and 61% of interactions, highlighting the uneven distribution of interactions per user.

At the highest level of pruning examined in our experiments, 100-core, a singular dataset (Gowalla) retained more than 1% of the users compared to the 0-coreset, while simultaneously retaining almost 40% of interactions and 54% of items. Whereas 100-core pruning may not be a common design choice, it highlights how far a dataset can be pruned while still retaining almost half of the interactions and items.

Data Pruning is acknowledged as a process that can reintroduce bias [22, 54], even if the original dataset was bias-free. Comparing algorithm performance on a dataset heavily altered, for example, by 50-core or 100-core pruning, would inevitably lead to results skewed in favour of high-activity users. Recommendations for cold-start users are known to be challenging [30], but a comparison of algorithm performance on these users is typically avoided by using pruned datasets. The impact of including cold-start users in evaluation scenarios will be discussed in more detail in the second part of this chapter.

In summary, we analysed the impact of data pruning even at low levels on the structural and distributional characteristics. We found that while item-centred characteristics were affected, it was to a lesser extent than user-centred characteristics. Overall, datasets reacted to the same levels of pruning in varying intensities; for example, Yelp experienced an 86% loss of users compared to Gowalla's 25% loss when 5-core pruning was applied. We observed that both structural and distributional characteristics were substantially changed by data pruning.

### 6.1.2  Algorithm Performance

The preceding analysis of dataset characteristics provides essential context for interpreting variance in algorithmic performance. Understanding how progressive pruning altered structural and distributional properties clarifies the conditions under which the algorithms were evaluated. In the following section, we build on these findings to examine how altered dataset characteristics influenced the performance of the tested algorithms.

In our experiment, we employed two complementary evaluation strategies to examine algorithm performance on progressively pruned datasets. In Phase 1, algorithms were trained and tested on equally pruned train and test sets, reflecting the common practice of using pruned datasets throughout recommender system research. In Phase 2, the algorithms were trained on the same training sets, while evaluation was conducted on test sets derived from unpruned data. In the discussion that follows, we first examine the results from Phase 1 and then compare them with those from Phase 2, highlighting what these contrasts reveal about the implications of training and testing on pruned datasets.

*6.1.2.1      Phase 1: Training and Testing on Pruned Datasets*

According to Beel and Brunel [8], the key question when examining the effects of data pruning is whether pruning leads to changes in the performance ranking of algorithms. During Phase 1, we observed that the performance ranks of most algorithms varied across the progressive pruning levels. The Random algorithm represented the sole exception, ranking last across every coreset examined in Phase 1, with multiple 0-values observed for average nDCG@10.

PopScore, as the second baseline algorithm, displayed improvements for higher levels of pruning, especially on denser datasets, such as MovieLens. In one case, on the 50-core pruned MovieLens dataset, PopScore even outperformed all remaining algorithms. The dataset's inherent structure can likely explain this ranking. As shown in the previous section, MovieLens contains a small number of very popular items with which many users have interacted. Because PopScore recommends items based on popularity, its improved performance on extensively pruned data is not unexpected. Examining the characteristics of 50-core pruned MovieLens reveals conditions particularly favourable towards a popularity-based algorithm: both the item retention rate and the $Gini_{Item}$ were observed to be much higher for MovieLens than for the compared datasets after 50-core pruning.

In general, we observed rank changes to be more frequent and more extreme in two datasets: the MovieLens and Yelp datasets. The first indication that these datasets differed was that User KNN, which ranked undisputedly first on the remaining three datasets across all cores, only experienced rank changes on MovieLens and Yelp. Analysing the absolute nDCG@10 values for these datasets, however, paints a different picture: Even the Yelp dataset's progressive pruning led to an increase in absolute nDCG@10 values with every core level, whereas the performance on the MovieLens dataset declined with each successive core level. On Yelp pruned to cores 5 to 20, User KNN consistently ranked second, supported by a steady rise in nDCG@10 scores per core, and was surpassed only by Implicit MF, which achieved its peak performance on these specific coresets in Phase 1.

Structural characteristics of the MovieLens coresets account for the rank changes observed between the two neighbourhood-based approaches, Item KNN and User KNN. For the unpruned dataset (0-core) and the heavily pruned 100-coreset, User KNN outperformed Item KNN, whereas the medium levels of pruning favoured Item KNN. This observation can be attributed to the structural characteristics observed when pruning the MovieLens dataset: the long-tail item distribution in the unpruned version makes recommending based on item similarity difficult, as many items receive very few interactions, leading to better performance for User KNN over Item KNN. Because the 100-coreset retains only users with extensive interaction histories, it provides ideal conditions for a user-similarity algorithm and therefore, explains User KNN's advantage on this coreset.

The Matrix Factorisation-based algorithms, Biased MF and Implicit MF, both increased in performance scores as pruning levels rose, exhibiting the most significant performance improvement compared to the 0-core datasets on average on the 100-coreset. This behaviour is not surprising, as it has been acknowledged that algorithms employing Matrix Factorisation display performance improvements as Sparsity decreases [46]. As we have shown in the analysis of the coreset characteristics, Density increased with each level of pruning applied, thus Sparsity ($Sparsity = 1 - Density$) decreased and performance for Biased MF and Implicit MF improved.

BPR profited to an extent of rising density by progressive pruning, exhibiting enhanced performance scores with each core-level for Gowalla and both Amazon datasets, though to a lesser degree for Amazon Toys. The performance on the denser datasets MovieLens and Yelp dropped sharply at 100-core pruning. For example, BPR's score declined by 44% (from 0.025 to 0.014) compared to the 50-core level when trained and tested on the 100-coreset. This decline aligns with the sharp reduction in interactions both datasets experienced when pruned to 100 cores, as neither retained more than 15% of interactions at this pruning level. Interaction loss poses a disadvantage for algorithms like BPR, which depend on positive-negative item pairs to learn preferences [37, 40].

Overall, we observed positive trends for the group of traditional algorithms under progressively increased pruning conditions. The group containing the modern algorithms displayed less homogeneous behaviour across pruning levels, with individual algorithms reacting differently to data reduction. For example, DiffRec responded to low pruning levels with an average initial performance improvement of 25% (5-core) and 15% (10-core) compared to the unpruned baseline. Further pruning continuously diminished this effect to the point of 100-core pruning, exhibiting an average performance decline of 80%.

MultiVAE exhibited a similar trend to DiffRec, with a decrease in performance scores across higher pruning levels in response to increased pruning intensity. Compared to the rest of the modern algorithms, MultiVAE reacted most sensitively to even low-level pruning, showing performance declines of 22% and 27% at the 5-core and 10-core levels, respectively. We observed that MultiVAE ranked higher on the unpruned versions of Yelp and MovieLens, with a sharp decline on Yelp as early as when 5-core pruning was applied. This correlates with the data loss Yelp experienced when 5-core pruning was applied: relative user retention measured 14%, while interaction retention dropped to 58%. MultiVAE achieved higher performance scores on the dense datasets; however, it exhibited a drastic decline in performance with data reduction.

DMF appeared less sensitive to data reduction, with only minor changes in performance compared to the unpruned baseline; however, the MovieLens dataset elicited steeper declines across pruning levels. Although the highest score on the MovieLens coresets was obtained when training and testing on the unpruned version, a comparison of the 50-core and 100-core results showed a performance improvement. This trend was observed for all examined datasets and resembled the behaviour of the traditional Matrix Factorisation algorithms. Overall, DMF improved on average by 48% on the 100-coresets.

SimpleX responded differently to 100-core pruning for the MovieLens and Yelp datasets than DMF. For MovieLens, a steady decline was observed across core levels. SimpleX achieved the highest performance score on Yelp when 50-core pruning was applied. When 100-core pruned, the Amazon datasets and Gowalla yielded the highest nDCG@10 values for SimpleX. This division between the two dataset groups, contrasting in density and shape, resulting in differing scores for algorithms, was a recurring observation throughout the first phase of the experiment.

To summarise, we observed the algorithm groups to react differently to data pruning, resulting in various rank changes both for algorithms performing better on entire datasets as well as rank changes across pruning levels for a singular dataset. The baseline algorithms improved towards the higher levels of pruning, while the modern algorithms tended to react more sensitively to data reduction. This aligns with the commonly accepted assumption that deep learning algorithms perform better when the amount of data increases [45]. In contrast, the traditional algorithms showed an overall performance improvement with each level of pruning applied. Additionally, we observed that the traditional algorithms, on average, outperformed the group of modern algorithms across all core levels.

*6.1.2.2        Phase 2: Training on Pruned Datasets, Testing on Unpruned Datasets*
We demonstrated, through the examination of Phase 1, that training and evaluating algorithms on pruned versions of datasets can lead to performance rank changes between algorithms, both across different pruning levels and across entire datasets. In Phase 2, we trained the algorithms on the same training sets as in Phase 1, but the evaluation was performed on test sets derived from an unpruned dataset. To enable a fair comparison, these test sets had the same size as the pruned test sets from Phase 1, but included low-activity users. Furthermore, it was ensured that the test sets only included unseen data by choosing disjoint subsets of the training sets.

The baseline algorithms, PopScore and Random, exhibited notably different behaviour compared to Phase 1, when evaluation included recommendations for low-activity users. For instance, PopScore's overall positive trend, especially pronounced at core levels 20 and above, flattened almost entirely, showing only a slight increase for the 20-core pruned MovieLens dataset. Random fared even worse: on three datasets (both Amazon and Gowalla), it failed to produce a single non-zero average nDCG@10 value.

The majority of traditional algorithms (80%) experienced a similar trend as the baseline algorithms in Phase 2. While there were minimal differences in average nDCG@10 for the datasets, the overall trend was the same: the algorithms only improved their performance on the 20-core pruned MovieLens training set. Some of the algorithms in this group showed slight nDCG@10 increases on individual coresets when compared to the lowest core-level (5-core), but the absolute differences never exceeded 0.0007 and are thereby deemed insignificant. Compared to the performance scores achieved in Phase 1 on the pruned test sets, the traditional algorithms (Biased MF, Implicit MF, Item KNN, User KNN) suffered a decline of an average 95% across the pruning levels.

The only exception to this decline was BPR, which overall performed better under the evaluation conditions in Phase 2 and displayed an average increase in nDCG@10 of 143% across all pruned training sets when evaluated on unpruned test sets. Compared to Phase 1, BPR's improvement in the performance-based ranking of algorithms was only second to SimpleX: on the entirety of all coresets for both Amazon datasets and the Gowalla datasets, BPR ranked in second place in Phase 2. The lowest rank we observed was for the MovieLens datasets, when pruned to 10 cores and higher, where BPR ranked 4$^{th}$. For the highly challenging 100-core pruned Yelp dataset, BPR performed best of all algorithms, displaying a stronger tolerance towards heavy data reductions than the remaining algorithms. While BPR did not respond to data reduction as strongly, comparing the nDCG@10 values across the pruning levels in Phase 2 still showed a decrease as the pruning intensity increased. Applying 10-core pruning (0.051) lowered the average nDCG@10 by 19% compared to 5-core pruning (0.063), and 100-core pruning caused an even steeper decline of 78%, down to 0.014. The steady decline shows that BPR's performance was progressively hindered as the pruning levels increased.

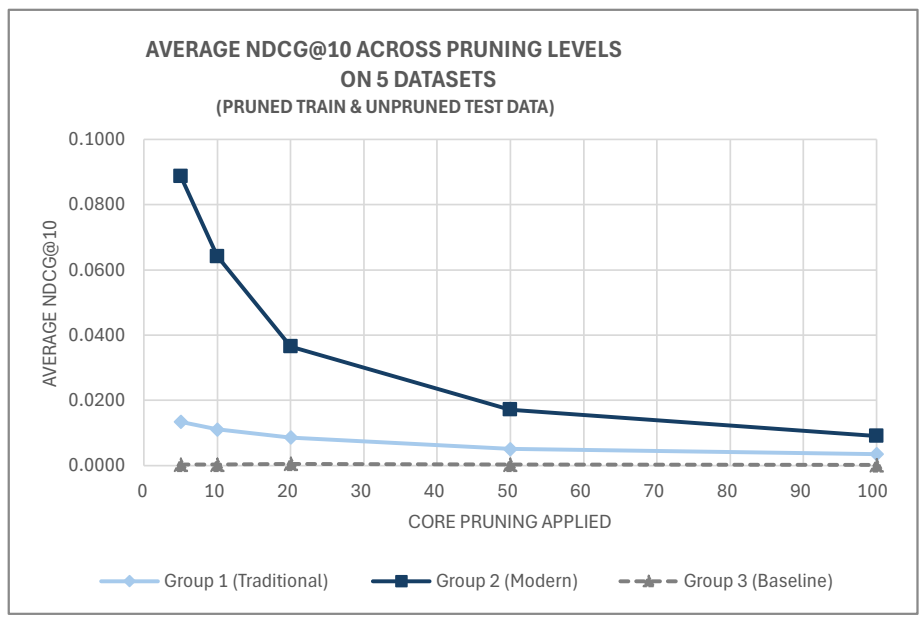

*Figure 14 Figure X – Average nDCG@10 Across Pruning Levels with Pruned Training and Unpruned Test Data:*
*Average nDCG@10 for traditional, modern, and baseline algorithms across five datasets when models are trained on pruned datasets but evaluated on unpruned test sets.*

The modern algorithms achieved much higher absolute performance scores under Phase 2 evaluation compared to the traditional algorithms. Across all examined coresets, the traditional algorithms attained a mean nDCG@10 of 0.010, while the modern algorithms achieved a mean of 0.038. All modern algorithms exhibited a declining performance trend as pruning levels increased (**Figure 14**), though their absolute nDCG@10 in some cases were higher than the corresponding Phase 1 values. For example, DiffRec achieved values for 5- to 20-core pruning, which were an average 370% higher than the scores for the corresponding training sets in Phase 1. However, increasing the pruning level from 5-core to 10-core resulted in a 39% drop in mean nDCG@10 (0.116 to 0.071), and 100-core pruning even dropped by 98% to 0.002.

It is worth mentioning that training SimpleX on pruned and testing it on unpruned data yielded the highest observed nDCG@10 scores in the entire experiment. On both Amazon and the Gowalla datasets, SimpleX was the only algorithm we observed to reach mean nDCG@10 values above 0.25, with a maximum score attained on the Amazon Toys dataset pruned at core-level 5 (0.49). Nonetheless, a declining trend in these scores was observed as pruning intensified, which demonstrates the detrimental impact that progressive data pruning had on the modern algorithms examined in this thesis.

Overall, the findings contrasted sharply with those of Phase 1, as pruning the datasets did not yield any improvement in mean performance for either algorithm group. This aligns with the claim made by Beel and Brunel [8] in their initial study: Data Pruning artificially inflates the performance score when evaluated on test sets that have also undergone pruning. Algorithms previously experiencing upward trends on the pruned test sets did not display the same improvements on the unpruned test sets. These findings suggest that the improved performance on progressively pruned datasets stems from the test set composition, as the test sets were the only element that differed between the two evaluation settings.

The test sets in Phase 1 were derived by splitting the pruned core datasets into a train and test set, to resemble the usage of a ready-to-use pruned dataset (e.g. most of the MovieLens datasets). If algorithm evaluation is conducted on such a dataset, the test set will underly similarly altered structural and distributional characteristics as the dataset it was split off from. As we have shown in the analysis of the dataset characteristics, pruning users according to their number of interactions in a dataset can lead to coresets that differ substantially in composition and structural characteristics.

Taken together, these observations allow us to return to the core research question and examine it in light of the results.

RQ: *How does progressive dataset pruning by increasing core thresholds affect the performance of recommender systems, the structural and distributional characteristics of the dataset?*

In order to respond to the first part of the question, it is necessary to consider the second part first. We have examined a variety of structural and distributional characteristics across pruned coresets and found all analysed characteristics to be altered by progressively pruning, though not to the same degree. Pruning users according to their total number of interactions in the datasets had a decreasing effect on the total number of items and interactions in the pruned coresets, and consequently, on all characteristics that depend on at least one of these total numbers. The continuous decrease in these characteristics with rising pruning levels was unsurprising: what stood out was the variability in dataset responses.

Structural and distributional characteristics were affected unevenly by pruning levels, but the clearest difference appeared in the retention of users, items and interactions at each stage (**Figure 13**). For instance, as early as just 5-core pruning, the gap between the highest user retention (Gowalla) and the lowest (Yelp) exceeded 60 percentage points. This consequently resulted in vastly different compositions for the core-pruned datasets, as some datasets retained much higher rates of their interactions than others. Even though all datasets had the same number of interactions in the 0-coresets, merely 5-core pruning led to almost halving the dataset size for individual datasets (Yelp), while nearly retaining the full number of interactions (Gowalla, MovieLens).

Our results additionally highlighted that the user retention rates dropped drastically for commonly chosen core-levels of 5, 10 and 20: only a single dataset retained more than 30% of users in its 20-coreset, while the majority of datasets only contained 7-15% of users compared to their unpruned versions. These varying degrees of user retention were also visible in the distributional characteristics when comparing the $Gini_{User}$. Especially dense datasets with an extremely short head, long tail distribution of interactions, such as the MovieLens dataset, displayed a trend toward uniformity across pruning levels in $\text{Gini}_{\text{User}}$. Whereas $\text{Gini}_{\text{Item}}$ remained considerably high, leaning towards an unequal distribution of interactions across items.

In summary, progressive pruning altered both structural and distributional characteristics across all datasets; however, the extent of these changes was strongly influenced by the inherent composition of the unpruned (0-core) datasets.

Dataset characteristics, such as density, shape, and the average number of interactions per user/item, are known to influence algorithm performance [18]. Therefore, the fluctuations observed in our experiments when comparing algorithm performance across datasets and their core-pruned versions were expected after analysing the effect pruning had on the dataset characteristics. Nonetheless, some algorithms showed distinctly different behaviour on individual coresets, a pattern we hypothesise arises from their underlying operating mechanisms. For example, Matrix Factorisation-based algorithms tend to improve performance as data sparsity decreases, a trend mirrored in the results of our experimentation in Phase 1 and the dataset analysis.

The key question of whether progressive data pruning leads to changes in the performance ranking of algorithms was answered affirmatively for all examined datasets. Rank changes occurred more frequently and with greater magnitude on the two datasets with the highest density, largest values for shape and average number of interactions per item: MovieLens and Yelp. This observation remained true for both Phase 1 and Phase 2.

In Phase 1, overall trends emerged across the three algorithm groups: traditional algorithms responded positively to higher pruning levels, whereas modern algorithms tended to show declining performance as pruning intensified. In Phase 2 – where test sets still included low-activity users while training was performed on pruned data – the performance rankings changed markedly, and traditional algorithms experienced a substantial drop in performance. This suggests that the improvements observed for traditional algorithms in Phase 1 were largely due to the progressively easier test sets, as higher pruning levels increasingly retained only high-activity users in the coresets.

The superior Phase 2 performance of modern algorithms supports the assumption that they are better equipped than traditional algorithms to handle typical recommender system challenges, such as data sparsity and the user cold-start problem [2, 31]. A scenario with cold-start users was simulated in Phase 2 by including low-activity users in the test sets.

To revisit the core research question: **progressive pruning altered all examined dataset characteristics**, yet the underlying composition of the original datasets remained reflected in these altered characteristics. Both the inherent dataset composition and the pruning-induced changes influenced algorithm performance, confirming that **data pruning affects performance ranking**. However, rank changes cannot be explained solely by the progressively pruned training sets, as the composition of the test sets and the intrinsic mechanisms of the algorithms also play significant roles.

## 7. Conclusion

This thesis set out to analyse the impact of progressive data pruning on dataset characteristics and algorithm performance. To this end, we first analysed five datasets in both unpruned and progressively pruned states (5, 10, 20, 50, and 100-core pruning) to evaluate how their characteristics evolved under increasing pruning intensity.

The results show that inherent dataset characteristics observed in the unpruned versions shape the extent to which they are altered by progressive pruning. For example, datasets with a high average number of interactions per user retained a much larger proportion of their original user base as pruning intensified. In contrast, datasets with lower average interactions per user, e.g. Yelp, quickly experienced heavy user loss and preserved fewer total interactions than datasets with higher interaction averages per user. Consequently, the resulting coresets varied significantly in terms of characteristics, total size, and user retention. For example, when 5-core pruning was applied, Gowalla retained as many as 75% of its users and 98% of its interactions, whereas Yelp lost 86% of its users and only retained 58% of its interactions. This contrast widened at higher pruning levels: under 100-core pruning on Yelp retained merely 0.12% of its original users, while Gowalla still maintained about 5%.

When datasets were pruned at the commonly used 20-core threshold, 80% of them retained 15% or fewer of their original users. This pruning strategy resembles the procedure applied to the popular MovieLens datasets, which are widely regarded as benchmark data for evaluating recommender systems. Using datasets pruned in this severity means that roughly 85% of users are ignored when evaluating an algorithm, thereby excluding the very group – low-activity or cold-start users – that is generally considered challenging for generating suitable recommendations.

Pruned datasets differ substantially from data encountered in real-world settings, as they primarily contain users with comparatively rich interaction histories. Consequently, good results achieved on such datasets cannot necessarily be generalised to other scenarios. Furthermore, pruning can introduce bias into datasets, in our case by skewing performance towards high-activity users. This raises concerns about fairness in recommender systems and has become the subject of recent research.

The influence of pruning on performance scores and, ultimately, the performance ranking of algorithms was assessed by training and testing eleven algorithms on increasingly pruned datasets. The selected algorithms – covering baseline, traditional, and modern methods – allowed us to analyse whether pruning affected these groups to different extents.

We found that specific datasets as a whole, as well as particular coresets, provided favourable conditions for individual algorithms. For example, denser datasets benefited matrix factorisation-based methods, while coresets with few low-activity items favoured Item KNN. Overall, the results suggest that traditional algorithms tended to improve as pruning levels increased, whereas modern algorithms showed no corresponding performance increase on progressively pruned coresets. In summary, changes in the performance rank of algorithms were observed across all datasets, with a higher frequency and greater magnitude on the denser datasets, MovieLens and Yelp.

To determine whether the "easier" test sets – those in which low-activity users are successively removed, leaving only high-activity users – contributed to the positive trend observed for some algorithms across pruning levels, we conducted a second evaluation. This evaluation utilised the same training sets as before, but the test sets were derived from the unpruned datasets, thereby reintroducing low-activity users. While rank changes occurred across coresets and entire datasets under these evaluation conditions as well, the overall trends and performance values declined sharply, particularly for the traditional algorithms. Whereas traditional algorithms had outperformed modern ones across most coresets in the first evaluation phase, testing on unpruned datasets yielded higher overall performance scores for the modern algorithms. A comparison of score evolution across increasing pruning levels showed an overall decline in performance as pruning intensified for all examined algorithms of both groups.

Combining our findings from the dataset-characteristic analysis, the experiments with training and testing on pruned data, and the experiments with training on pruned data while testing on unpruned data, we conclude that data pruning can artificially inflate performance scores. This effect occurs when evaluation is carried out on test sets pruned to the same level as the training data. Furthermore, specific datasets and pruning levels favoured individual algorithms, indicating that evaluating only on single datasets or individual coresets is not advisable for a balanced comparison of algorithm performance. Results based on such narrow evaluations cannot necessarily be generalised to other scenarios that involve different datasets. One such scenario is the data a recommender system encounters in real-world use. Choosing an unsuitable algorithm in this context could lead to performance that differs significantly from what offline evaluations suggested.

Besides failing to deliver generally applicable results, data pruning also hampers effective comparisons between algorithms in research. Selecting a coreset that is especially favourable to a particular algorithm can skew performance results in its favour, while making other comparable algorithms appear less suitable.

Data pruning can help reduce computational load and training time, thereby supporting greener recommender-system research, provided acceptable trade-offs between data reduction and performance are possible. However, we concur with evaluation frameworks that recommend using pruned datasets only as a supplement to unpruned datasets. Alternatively, training on pruned data while evaluating on unpruned data could offer a compromise that lowers training time and computational cost while still testing on a heterogeneous group of users. Additionally, our results highlight the importance of clearly documenting all preprocessing steps applied to datasets in recommender-systems research. Although many studies apply data pruning during pre-processing or use pre-pruned datasets, such as MovieLens, crucial documentation of dataset choice and preprocessing steps, including data pruning, is frequently absent. Without precise knowledge of a dataset's preprocessing history, replicating experiments and comparing algorithms under equal conditions becomes nearly impossible.

In conclusion, this thesis demonstrates that progressive data pruning can strongly influence both dataset composition and algorithm evaluation, leading to inflated performance scores and limited generalisability if not handled with care. By combining detailed dataset analyses with extensive algorithmic assessment, we provide evidence that pruning should be applied only as a complement to unpruned data and documented transparently. Future recommender-system research can build on these insights by developing frameworks and preprocessing standards that balance computational efficiency with fairness and real-world applicability.

## 8. Future Work and Limitations

Due to computational limitations encountered during the experiment, several concessions had to be made to facilitate the evaluation of 11 algorithms on five datasets.

**Hyperparameter-Optimisation**

All models were trained with their default parameter values, respectively set by either the LensKit or the RecBole library, as the focus of this study was not to achieve the highest possible performance for all algorithms but to compare how the increasingly pruned versions of the datasets affected the performance of the models relatively. Future work should consider implementing hyperparameter-tuning.

**K-Fold Cross Validation**

Due to the extensive computation times, it was not possible to include K-Fold Cross-Validation in this study. Future research could consider implementing cross-validation with k = 5 or k = 10.

**Algorithm Selection**

To further reduce the computational cost, the deep learning models investigated in this study were chosen for their lower computational complexity and shorter training times. It would be of interest to see how more complex deep learning models are impacted by data pruning.

**Metrics**

Our research focused on ranking-based performance metrics, researchers could examine the impact data pruning has on, e.g. beyond-accuracy metrics for deep learning models.

**Datasets**

Due to the scope of this work, only five datasets could be included in the experiments; further research should focus on a more diverse variety of data sources.

**User-Based and Item-Based Pruning**

In our experiment, we pruned interactions based on the total number of interactions the corresponding user had in the dataset. It is interesting to examine how pruning based on item interactions, e.g. pruning all interactions with an item if it is interacted with less than a specific number of times, impacts both dataset characteristics and algorithm performance.

Addressing these limitations in future research will not only strengthen the validity and generalisability of our findings but also provide a richer foundation for developing fair, efficient, and reproducible recommender-system evaluations. By building on the proposed directions, subsequent studies can refine methods, broaden datasets, and deepen understanding of how data pruning shapes both dataset characteristics and algorithm performance.

# 9. Summary

This thesis investigates how data pruning – removing users whose total number of interactions falls below a defined threshold – affects both datasets' characteristics and algorithm performance in recommender systems. Five widely used datasets were examined in their unpruned form and at 5-, 10-, 20-, 50-, and 100-core pruning levels. For each stage, we analysed structural characteristics, such as density, shape, and the average number of interactions per user/item, as well as distributional characteristics measured by Gini coefficients for users and items. Retention rates of users, items, and interactions were tracked to reveal how dataset composition evolves as low-activity users are successively removed.

Dataset characteristics were found to be influenced by pruning to varying degrees. Even at the lowest pruning level of 5-core, the Yelp dataset retained only 14% of its users and 58% of its interactions. In contrast, Gowalla preserved 75% of its users and 98% of its interactions, illustrating how differently datasets can react to identical pruning thresholds. At the commonly used 20-core level, all but one dataset were reduced to 15% or fewer of their original users, and at the highest pruning level (100-core), Yelp retained a mere 0.12% of its user base. These findings highlight the heterogeneous impact of progressive pruning on structural and distributional characteristics across datasets.

Building on this analysis, algorithm performance was evaluated across the same progressive pruning levels in two complementary settings. Phase 1 involved training and testing on increasingly pruned datasets, while Phase 2 used the same pruned training sets but tested on unpruned datasets to reintroduce low-activity users to the evaluation. While traditional algorithms overall showed improving performance with rising pruning levels in Phase 1, this trend diminished significantly in Phase 2, where modern algorithms continuously outperformed the majority of traditional algorithms. The modern algorithms exhibited a general trend of declining performance as pruning levels increased during both phases, despite average nDCG@10 values being higher in Phase 2. Rank changes occurred between algorithms in both phases, across entire datasets and for specific coresets. On the denser datasets, Yelp and MovieLens, both frequency and magnitude of rank changes increased across core levels.

The two-phase evaluation revealed how pruning can artificially inflate performance scores and alter algorithm rankings when testing is conducted on equally pruned data, and how these effects change when testing occurs instead on unpruned datasets.

Together, these analyses of algorithm performance and dataset characteristics provide a detailed picture of how progressive data pruning influences both the statistical structure of recommender-systems datasets and the measured performance of baseline, traditional, and modern algorithms. The results highlight that pruning levels have a strong impact on dataset composition and algorithmic evaluation, and the careful selection of datasets and coresets can provide beneficial conditions for individual algorithms. The findings underline the importance of selecting diverse datasets, using core-pruned datasets only as supplementary analysis to unpruned datasets, and carefully documenting every preprocessing step to which datasets are subjected.

# 11. Appendix

| Phase 1 (Training & Testing on Pruned Datasets) nDCG@10 | | | | | | Phase 2 (Testing on Unpruned Datasets) nDCG@10 | | | | | |
|---|---|---|---|---|---|---|---|---|---|---|---|
| **Biased MF** | | | | | | **Biased MF** | | | | | |
| Core | Amazon CD | Amazon Toy | Gowalla | MovieLens | Yelp | Core | Amazon CD | Amazon Toy | Gowalla | MovieLens | Yelp |
| 0 | 0.0000 | 0.0000 | 0.0002 | 0.0070 | 0.0007 | | | | | | |
| 5 | 0.0000 | 0.0000 | 0.0002 | 0.0065 | 0.0007 | 5 | 0.0000 | 0.0000 | 0.0000 | 0.0004 | 0.0000 |
| 10 | 0.0000 | 0.0000 | 0.0003 | 0.0059 | 0.0006 | 10 | 0.0000 | 0.0000 | 0.0000 | 0.0003 | 0.0000 |
| 20 | 0.0000 | 0.0000 | 0.0004 | 0.0063 | 0.0008 | 20 | 0.0000 | 0.0000 | 0.0000 | 0.0010 | 0.0000 |
| 50 | 0.0006 | 0.0000 | 0.0008 | 0.0070 | 0.0013 | 50 | 0.0000 | 0.0000 | 0.0000 | 0.0003 | 0.0000 |
| 100 | 0.0008 | 0.0000 | 0.0013 | 0.0097 | 0.0022 | 100 | 0.0000 | 0.0000 | 0.0000 | 0.0003 | 0.0000 |
| Min | 0.0000 | 0.0000 | 0.0002 | 0.0059 | 0.0006 | Min | 0.0000 | 0.0000 | 0.0000 | 0.0003 | 0.0000 |
| Max | 0.0008 | 0.0000 | 0.0013 | 0.0097 | 0.0022 | Max | 0.0000 | 0.0000 | 0.0000 | 0.0010 | 0.0000 |
| Average | 0.0003 | 0.0000 | 0.0005 | 0.0071 | 0.0011 | Average | 0.0000 | 0.0000 | 0.0000 | 0.0005 | 0.0000 |
| **Implicit MF** | | | | | | **ImplicitMF** | | | | | |
| Core | Amazon CD | Amazon Toy | Gowalla | MovieLens | Yelp | Core | Amazon CD | Amazon Toy | Gowalla | MovieLens | Yelp |
| 0 | 0.0204 | 0.0021 | 0.0282 | 0.0424 | 0.0286 | | | | | | |
| 5 | 0.0202 | 0.0018 | 0.0298 | 0.0380 | 0.0332 | 5 | 0.0010 | 0.0001 | 0.0017 | 0.0021 | 0.0017 |
| 10 | 0.0215 | 0.0024 | 0.0328 | 0.0310 | 0.0366 | 10 | 0.0012 | 0.0001 | 0.0019 | 0.0017 | 0.0019 |
| 20 | 0.0266 | 0.0029 | 0.0398 | 0.0271 | 0.0452 | 20 | 0.0013 | 0.0002 | 0.0023 | 0.0042 | 0.0021 |
| 50 | 0.0410 | 0.0044 | 0.0555 | 0.0234 | 0.0606 | 50 | 0.0015 | 0.0002 | 0.0028 | 0.0011 | 0.0016 |
| 100 | 0.0565 | 0.0068 | 0.0789 | 0.0210 | 0.0751 | 100 | 0.0021 | 0.0002 | 0.0029 | 0.0004 | 0.0012 |
| Min | 0.0202 | 0.0018 | 0.0282 | 0.0210 | 0.0286 | Min | 0.0010 | 0.0001 | 0.0017 | 0.0004 | 0.0012 |
| Max | 0.0565 | 0.0068 | 0.0789 | 0.0424 | 0.0751 | Max | 0.0021 | 0.0002 | 0.0029 | 0.0042 | 0.0021 |
| Average | 0.0310 | 0.0034 | 0.0442 | 0.0305 | 0.0465 | Average | 0.0014 | 0.0002 | 0.0023 | 0.0019 | 0.0017 |
| **PopScore** | | | | | | **PopScore** | | | | | |
| Core | Amazon CD | Amazon Toy | Gowalla | MovieLens | Yelp | Core | Amazon CD | Amazon Toy | Gowalla | MovieLens | Yelp |
| 0 | 0.0057 | 0.0007 | 0.0056 | 0.0396 | 0.0075 | | | | | | |
| 5 | 0.0047 | 0.0007 | 0.0060 | 0.0353 | 0.0069 | 5 | 0.0002 | 0.0000 | 0.0004 | 0.0019 | 0.0004 |
| 10 | 0.0050 | 0.0010 | 0.0067 | 0.0298 | 0.0062 | 10 | 0.0002 | 0.0001 | 0.0004 | 0.0017 | 0.0003 |
| 20 | 0.0063 | 0.0009 | 0.0078 | 0.0272 | 0.0071 | 20 | 0.0003 | 0.0000 | 0.0005 | 0.0042 | 0.0004 |
| 50 | 0.0092 | 0.0014 | 0.0120 | 0.0251 | 0.0095 | 50 | 0.0004 | 0.0001 | 0.0006 | 0.0011 | 0.0004 |
| 100 | 0.0149 | 0.0019 | 0.0154 | 0.0278 | 0.0161 | 100 | 0.0004 | 0.0000 | 0.0004 | 0.0004 | 0.0004 |
| Min | 0.0047 | 0.0007 | 0.0056 | 0.0251 | 0.0062 | Min | 0.0002 | 0.0000 | 0.0004 | 0.0004 | 0.0003 |
| Max | 0.0149 | 0.0019 | 0.0154 | 0.0396 | 0.0161 | Max | 0.0004 | 0.0001 | 0.0006 | 0.0042 | 0.0004 |
| Average | 0.0076 | 0.0011 | 0.0089 | 0.0308 | 0.0089 | Average | 0.0003 | 0.0000 | 0.0005 | 0.0018 | 0.0004 |
| **User KNN** | | | | | | **User KNN** | | | | | |
| Core | Amazon CD | Amazon Toy | Gowalla | MovieLens | Yelp | Core | Amazon CD | Amazon Toy | Gowalla | MovieLens | Yelp |
| 0 | 0.0337 | 0.0082 | 0.0512 | 0.0504 | 0.0312 | | | | | | |
| 5 | 0.0341 | 0.0081 | 0.0523 | 0.0428 | 0.0317 | 5 | 0.0018 | 0.0004 | 0.0031 | 0.0022 | 0.0017 |
| 10 | 0.0364 | 0.0083 | 0.0546 | 0.0340 | 0.0332 | 10 | 0.0019 | 0.0004 | 0.0030 | 0.0017 | 0.0018 |
| 20 | 0.0450 | 0.0092 | 0.0621 | 0.0279 | 0.0419 | 20 | 0.0023 | 0.0005 | 0.0036 | 0.0043 | 0.0021 |
| 50 | 0.0621 | 0.0095 | 0.0795 | 0.0239 | 0.0628 | 50 | 0.0026 | 0.0005 | 0.0040 | 0.0009 | 0.0015 |
| 100 | 0.0820 | 0.0103 | 0.1013 | 0.0251 | 0.0780 | 100 | 0.0026 | 0.0003 | 0.0036 | 0.0006 | 0.0014 |
| Min | 0.0337 | 0.0081 | 0.0512 | 0.0239 | 0.0312 | Min | 0.0018 | 0.0003 | 0.0030 | 0.0006 | 0.0014 |
| Max | 0.0820 | 0.0103 | 0.1013 | 0.0504 | 0.0780 | Max | 0.0026 | 0.0005 | 0.0040 | 0.0043 | 0.0021 |
| Average | 0.0489 | 0.0089 | 0.0668 | 0.0340 | 0.0464 | Average | 0.0022 | 0.0004 | 0.0034 | 0.0019 | 0.0017 |
| **Item KNN** | | | | | | **Item KNN** | | | | | |
| Core | Amazon CD | Amazon Toy | Gowalla | MovieLens | Yelp | Core | Amazon CD | Amazon Toy | Gowalla | MovieLens | Yelp |
| 0 | 0.0107 | 0.0030 | 0.0216 | 0.0448 | 0.0066 | | | | | | |
| 5 | 0.0130 | 0.0037 | 0.0223 | 0.0440 | 0.0102 | 5 | 0.0008 | 0.0002 | 0.0014 | 0.0023 | 0.0006 |
| 10 | 0.0176 | 0.0041 | 0.0246 | 0.0371 | 0.0146 | 10 | 0.0010 | 0.0003 | 0.0014 | 0.0020 | 0.0008 |
| 20 | 0.0245 | 0.0053 | 0.0296 | 0.0314 | 0.0227 | 20 | 0.0012 | 0.0003 | 0.0017 | 0.0049 | 0.0011 |
| 50 | 0.0331 | 0.0060 | 0.0379 | 0.0246 | 0.0229 | 50 | 0.0014 | 0.0002 | 0.0019 | 0.0008 | 0.0005 |
| 100 | 0.0413 | 0.0068 | 0.0424 | 0.0145 | 0.0161 | 100 | 0.0015 | 0.0002 | 0.0019 | 0.0003 | 0.0004 |
| Min | 0.0107 | 0.0030 | 0.0216 | 0.0145 | 0.0066 | Min | 0.0008 | 0.0002 | 0.0014 | 0.0003 | 0.0004 |
| Max | 0.0413 | 0.0068 | 0.0424 | 0.0448 | 0.0229 | Max | 0.0015 | 0.0003 | 0.0019 | 0.0049 | 0.0011 |
| Average | 0.0234 | 0.0048 | 0.0297 | 0.0327 | 0.0155 | Average | 0.0012 | 0.0003 | 0.0017 | 0.0021 | 0.0007 |
| **BPR** | | | | | | **BPR** | | | | | |
| Core | Amazon CD | Amazon Toy | Gowalla | MovieLens | Yelp | Core | Amazon CD | Amazon Toy | Gowalla | MovieLens | Yelp |
| 0 | 0.0050 | 0.0003 | 0.0146 | 0.0367 | 0.0132 | | | | | | |
| 5 | 0.0050 | 0.0003 | 0.0150 | 0.0350 | 0.0150 | 5 | 0.0448 | 0.2004 | 0.0230 | 0.0319 | 0.0141 |
| 10 | 0.0064 | 0.0005 | 0.0180 | 0.0304 | 0.0172 | 10 | 0.0324 | 0.1655 | 0.0211 | 0.0249 | 0.0106 |
| 20 | 0.0091 | 0.0005 | 0.0216 | 0.0273 | 0.0229 | 20 | 0.0211 | 0.1234 | 0.0183 | 0.0131 | 0.0062 |
| 50 | 0.0151 | 0.0012 | 0.0296 | 0.0243 | 0.0248 | 50 | 0.0147 | 0.0702 | 0.0103 | 0.0075 | 0.0036 |
| 100 | 0.0264 | 0.0025 | 0.0401 | 0.0138 | 0.0019 | 100 | 0.0117 | 0.0417 | 0.0068 | 0.0027 | 0.0049 |
| Min | 0.0050 | 0.0003 | 0.0146 | 0.0138 | 0.0019 | Min | 0.0117 | 0.0417 | 0.0068 | 0.0027 | 0.0036 |
| Max | 0.0264 | 0.0025 | 0.0401 | 0.0367 | 0.0248 | Max | 0.0448 | 0.2004 | 0.0230 | 0.0319 | 0.0141 |
| Average | 0.0112 | 0.0009 | 0.0232 | 0.0279 | 0.0158 | Average | 0.0249 | 0.1202 | 0.0159 | 0.0160 | 0.0079 |

| Phase 1 (Training & Testing on Pruned Datasets) nDCG@10 | | | | | | Phase 2 (Testing on Unpruned Datasets) nDCG@10 | | | | | |
|---|---|---|---|---|---|---|---|---|---|---|---|
| **DiffRec** | | | | | | **DiffRec** | | | | | |
| **Core** | Amazon CD | Amazon Toy | Gowalla | MovieLens | Yelp | **Core** | Amazon CD | Amazon Toy | Gowalla | MovieLens | Yelp |
| 0 | 0.0078 | 0.0000 | 0.0097 | 0.0329 | 0.0108 | | | | | | |
| 5 | 0.0050 | 0.0004 | 0.0105 | 0.0395 | 0.0212 | 5 | 0.0054 | 0.0004 | 0.0111 | 0.4842 | 0.0765 |
| 10 | 0.0050 | 0.0004 | 0.0094 | 0.0370 | 0.0191 | 10 | 0.0049 | 0.0001 | 0.0104 | 0.3142 | 0.0229 |
| 20 | 0.0046 | 0.0006 | 0.0069 | 0.0304 | 0.0196 | 20 | 0.0038 | 0.0002 | 0.0079 | 0.0766 | 0.0086 |
| 50 | 0.0014 | 0.0007 | 0.0015 | 0.0213 | 0.0042 | 50 | 0.0015 | 0.0000 | 0.0053 | 0.0125 | 0.0023 |
| 100 | 0.0023 | 0.0005 | 0.0029 | 0.0060 | 0.0005 | 100 | 0.0008 | 0.0000 | 0.0025 | 0.0065 | 0.0012 |
| **Min** | 0.0014 | 0.0000 | 0.0015 | 0.0060 | 0.0005 | **Min** | 0.0008 | 0.0000 | 0.0025 | 0.0065 | 0.0012 |
| **Max** | 0.0078 | 0.0007 | 0.0105 | 0.0395 | 0.0212 | **Max** | 0.0054 | 0.0004 | 0.0111 | 0.4842 | 0.0765 |
| **Average** | 0.0044 | 0.0004 | 0.0068 | 0.0279 | 0.0126 | **Average** | 0.0033 | 0.0001 | 0.0074 | 0.1788 | 0.0223 |
| **DMF** | | | | | | **DMF** | | | | | |
| **Core** | Amazon CD | Amazon Toy | Gowalla | MovieLens | Yelp | **Core** | Amazon CD | Amazon Toy | Gowalla | MovieLens | Yelp |
| 0 | 0.0016 | 0.0005 | 0.0045 | 0.0328 | 0.0063 | | | | | | |
| 5 | 0.0024 | 0.0006 | 0.0037 | 0.0285 | 0.0056 | 5 | 0.0018 | 0.0007 | 0.0046 | 0.0268 | 0.0052 |
| 10 | 0.0029 | 0.0009 | 0.0028 | 0.0274 | 0.0051 | 10 | 0.0028 | 0.0007 | 0.0030 | 0.0272 | 0.0030 |
| 20 | 0.0025 | 0.0008 | 0.0035 | 0.0229 | 0.0057 | 20 | 0.0013 | 0.0007 | 0.0063 | 0.0170 | 0.0029 |
| 50 | 0.0061 | 0.0011 | 0.0112 | 0.0204 | 0.0083 | 50 | 0.0022 | 0.0005 | 0.0047 | 0.0191 | 0.0031 |
| 100 | 0.0117 | 0.0013 | 0.0125 | 0.0280 | 0.0142 | 100 | 0.0022 | 0.0004 | 0.0037 | 0.0146 | 0.0031 |
| **Min** | 0.0016 | 0.0005 | 0.0028 | 0.0204 | 0.0051 | **Min** | 0.0013 | 0.0004 | 0.0030 | 0.0146 | 0.0029 |
| **Max** | 0.0117 | 0.0013 | 0.0125 | 0.0328 | 0.0142 | **Max** | 0.0028 | 0.0007 | 0.0063 | 0.0272 | 0.0052 |
| **Average** | 0.0045 | 0.0009 | 0.0064 | 0.0267 | 0.0075 | **Average** | 0.0021 | 0.0006 | 0.0045 | 0.0209 | 0.0035 |
| **MultiVAE** | | | | | | **MultiVAE** | | | | | |
| **Core** | Amazon CD | Amazon Toy | Gowalla | MovieLens | Yelp | **Core** | Amazon CD | Amazon Toy | Gowalla | MovieLens | Yelp |
| 0 | 0.0055 | 0.0003 | 0.0054 | 0.0391 | 0.0143 | | | | | | |
| 5 | 0.0037 | 0.0005 | 0.0060 | 0.0339 | 0.0063 | 5 | 0.0031 | 0.0000 | 0.0058 | 0.0351 | 0.0068 |
| 10 | 0.0035 | 0.0006 | 0.0066 | 0.0303 | 0.0064 | 10 | 0.0021 | 0.0000 | 0.0057 | 0.0313 | 0.0044 |
| 20 | 0.0036 | 0.0002 | 0.0063 | 0.0274 | 0.0032 | 20 | 0.0013 | 0.0000 | 0.0063 | 0.0304 | 0.0054 |
| 50 | 0.0055 | 0.0005 | 0.0045 | 0.0105 | 0.0114 | 50 | 0.0007 | 0.0000 | 0.0049 | 0.0123 | 0.0044 |
| 100 | 0.0058 | 0.0006 | 0.0127 | 0.0115 | 0.0020 | 100 | 0.0004 | 0.0000 | 0.0034 | 0.0132 | 0.0034 |
| **Min** | 0.0035 | 0.0002 | 0.0045 | 0.0105 | 0.0020 | **Min** | 0.0004 | 0.0000 | 0.0034 | 0.0123 | 0.0034 |
| **Max** | 0.0058 | 0.0006 | 0.0127 | 0.0391 | 0.0143 | **Max** | 0.0031 | 0.0000 | 0.0063 | 0.0351 | 0.0068 |
| **Average** | 0.0046 | 0.0005 | 0.0069 | 0.0255 | 0.0073 | **Average** | 0.0015 | 0.0000 | 0.0052 | 0.0245 | 0.0049 |
| **Random** | | | | | | **Random** | | | | | |
| **Core** | Amazon CD | Amazon Toy | Gowalla | MovieLens | Yelp | **Core** | Amazon CD | Amazon Toy | Gowalla | MovieLens | Yelp |
| 0 | 0.0000 | 0.0000 | 0.0000 | 0.0002 | 0.0000 | | | | | | |
| 5 | 0.0000 | 0.0000 | 0.0000 | 0.0002 | 0.0000 | 5 | 0.0000 | 0.0000 | 0.0000 | 0.0002 | 0.0001 |
| 10 | 0.0000 | 0.0000 | 0.0000 | 0.0002 | 0.0001 | 10 | 0.0000 | 0.0000 | 0.0000 | 0.0002 | 0.0000 |
| 20 | 0.0000 | 0.0000 | 0.0000 | 0.0003 | 0.0001 | 20 | 0.0000 | 0.0000 | 0.0000 | 0.0002 | 0.0000 |
| 50 | 0.0001 | 0.0000 | 0.0000 | 0.0004 | 0.0004 | 50 | 0.0000 | 0.0000 | 0.0000 | 0.0002 | 0.0001 |
| 100 | 0.0002 | 0.0000 | 0.0001 | 0.0011 | 0.0005 | 100 | 0.0000 | 0.0000 | 0.0000 | 0.0001 | 0.0000 |
| **Min** | 0.0000 | 0.0000 | 0.0000 | 0.0002 | 0.0000 | **Min** | 0.0000 | 0.0000 | 0.0000 | 0.0001 | 0.0000 |
| **Max** | 0.0002 | 0.0000 | 0.0001 | 0.0011 | 0.0005 | **Max** | 0.0000 | 0.0000 | 0.0000 | 0.0002 | 0.0001 |
| **Average** | 0.0001 | 0.0000 | 0.0000 | 0.0004 | 0.0002 | **Average** | 0.0000 | 0.0000 | 0.0000 | 0.0002 | 0.0000 |
| **SimpleX** | | | | | | **SimpleX** | | | | | |
| **Core** | Amazon CD | Amazon Toy | Gowalla | MovieLens | Yelp | **Core** | Amazon CD | Amazon Toy | Gowalla | MovieLens | Yelp |
| 0 | 0.0028 | 0.0004 | 0.0071 | 0.0233 | 0.0064 | | | | | | |
| 5 | 0.0028 | 0.0004 | 0.0082 | 0.0195 | 0.0078 | 5 | 0.3057 | 0.4901 | 0.2683 | 0.0228 | 0.0211 |
| 10 | 0.0030 | 0.0006 | 0.0079 | 0.0182 | 0.0098 | 10 | 0.2396 | 0.3387 | 0.2454 | 0.0140 | 0.0135 |
| 20 | 0.0054 | 0.0005 | 0.0098 | 0.0176 | 0.0132 | 20 | 0.1543 | 0.2083 | 0.1854 | 0.0044 | 0.0093 |
| 50 | 0.0103 | 0.0011 | 0.0117 | 0.0193 | 0.0200 | 50 | 0.0690 | 0.0927 | 0.0996 | 0.0024 | 0.0065 |
| 100 | 0.0164 | 0.0019 | 0.0179 | 0.0154 | 0.0059 | 100 | 0.0323 | 0.0455 | 0.0449 | 0.0005 | 0.0032 |
| **Min** | 0.0028 | 0.0004 | 0.0071 | 0.0154 | 0.0059 | **Min** | 0.0323 | 0.0455 | 0.0449 | 0.0005 | 0.0032 |
| **Max** | 0.0164 | 0.0019 | 0.0179 | 0.0233 | 0.0200 | **Max** | 0.3057 | 0.4901 | 0.2683 | 0.0228 | 0.0211 |
| **Average** | 0.0068 | 0.0008 | 0.0104 | 0.0189 | 0.0105 | **Average** | 0.1602 | 0.2351 | 0.1687 | 0.0088 | 0.0107 |

**PRECISION@10 (PHASE 1)**
**Amazon CD**

| Core | BiasedMF | ImplicitMF | PopScore | User-KNN | Item-KNN | BPR | DiffRec | DMF | MultiVAE | Random | SimpleX |
|---|---|---|---|---|---|---|---|---|---|---|---|
| 0 | 0.0000 | 0.0060 | 0.0016 | 0.0106 | 0.0046 | 0.0019 | 0.0028 | 0.0008 | 0.0015 | 0.0000 | 0.0010 |
| 5 | 0.0000 | 0.0083 | 0.0020 | 0.0143 | 0.0066 | 0.0024 | 0.0023 | 0.0010 | 0.0015 | 0.0000 | 0.0013 |
| 10 | 0.0000 | 0.0130 | 0.0032 | 0.0222 | 0.0117 | 0.0041 | 0.0031 | 0.0024 | 0.0022 | 0.0000 | 0.0018 |
| 20 | 0.0000 | 0.0199 | 0.0049 | 0.0345 | 0.0197 | 0.0072 | 0.0038 | 0.0024 | 0.0034 | 0.0000 | 0.0043 |
| 50 | 0.0009 | 0.0359 | 0.0085 | 0.0546 | 0.0308 | 0.0132 | 0.0014 | 0.0064 | 0.0055 | 0.0000 | 0.0092 |
| 100 | 0.0012 | 0.0518 | 0.0131 | 0.0741 | 0.0391 | 0.0246 | 0.0021 | 0.0115 | 0.0053 | 0.0002 | 0.0151 |

**Amazon Toys**

| Core | BiasedMF | ImplicitMF | PopScore | User-KNN | Item-KNN | BPR | DiffRec | DMF | MultiVAE | Random | SimpleX |
|---|---|---|---|---|---|---|---|---|---|---|---|
| 0 | 0.0000 | 0.0008 | 0.0002 | 0.0027 | 0.0013 | 0.0001 | 0.0000 | 0.0002 | 0.0001 | 0.0000 | 0.0002 |
| 5 | 0.0000 | 0.0009 | 0.0003 | 0.0036 | 0.0020 | 0.0002 | 0.0002 | 0.0003 | 0.0002 | 0.0000 | 0.0003 |
| 10 | 0.0000 | 0.0016 | 0.0005 | 0.0052 | 0.0029 | 0.0003 | 0.0002 | 0.0005 | 0.0003 | 0.0000 | 0.0004 |
| 20 | 0.0000 | 0.0023 | 0.0007 | 0.0071 | 0.0044 | 0.0004 | 0.0003 | 0.0006 | 0.0001 | 0.0000 | 0.0004 |
| 50 | 0.0000 | 0.0040 | 0.0012 | 0.0088 | 0.0057 | 0.0012 | 0.0007 | 0.0010 | 0.0005 | 0.0000 | 0.0010 |
| 100 | 0.0000 | 0.0065 | 0.0017 | 0.0094 | 0.0064 | 0.0025 | 0.0005 | 0.0011 | 0.0007 | 0.0001 | 0.0017 |

**Gowalla**

| Core | BiasedMF | ImplicitMF | PopScore | User-KNN | Item-KNN | BPR | DiffRec | DMF | MultiVAE | Random | SimpleX |
|---|---|---|---|---|---|---|---|---|---|---|---|
| 0 | 0.0002 | 0.0159 | 0.0035 | 0.0277 | 0.0125 | 0.0084 | 0.0060 | 0.0030 | 0.0033 | 0.0000 | 0.0045 |
| 5 | 0.0002 | 0.0174 | 0.0039 | 0.0300 | 0.0136 | 0.0090 | 0.0069 | 0.0025 | 0.0037 | 0.0000 | 0.0054 |
| 10 | 0.0002 | 0.0210 | 0.0047 | 0.0356 | 0.0169 | 0.0114 | 0.0062 | 0.0023 | 0.0046 | 0.0000 | 0.0058 |
| 20 | 0.0004 | 0.0299 | 0.0066 | 0.0478 | 0.0237 | 0.0158 | 0.0054 | 0.0032 | 0.0054 | 0.0000 | 0.0084 |
| 50 | 0.0007 | 0.0478 | 0.0106 | 0.0693 | 0.0349 | 0.0251 | 0.0015 | 0.0101 | 0.0042 | 0.0000 | 0.0112 |
| 100 | 0.0013 | 0.0695 | 0.0141 | 0.0894 | 0.0400 | 0.0352 | 0.0029 | 0.0127 | 0.0120 | 0.0001 | 0.0174 |

**MovieLens**

| Core | BiasedMF | ImplicitMF | PopScore | User-KNN | Item-KNN | BPR | DiffRec | DMF | MultiVAE | Random | SimpleX |
|---|---|---|---|---|---|---|---|---|---|---|---|
| 0 | 0.0020 | 0.0141 | 0.0130 | 0.0160 | 0.0149 | 0.0127 | 0.0120 | 0.0114 | 0.0137 | 0.0001 | 0.0089 |
| 5 | 0.0023 | 0.0156 | 0.0146 | 0.0174 | 0.0177 | 0.0147 | 0.0165 | 0.0123 | 0.0145 | 0.0001 | 0.0090 |
| 10 | 0.0028 | 0.0184 | 0.0175 | 0.0200 | 0.0215 | 0.0172 | 0.0210 | 0.0156 | 0.0173 | 0.0002 | 0.0110 |
| 20 | 0.0037 | 0.0212 | 0.0208 | 0.0219 | 0.0242 | 0.0206 | 0.0232 | 0.0174 | 0.0206 | 0.0003 | 0.0140 |
| 50 | 0.0056 | 0.0229 | 0.0245 | 0.0236 | 0.0244 | 0.0240 | 0.0210 | 0.0197 | 0.0098 | 0.0004 | 0.0191 |
| 100 | 0.0075 | 0.0219 | 0.0270 | 0.0249 | 0.0144 | 0.0134 | 0.0057 | 0.0268 | 0.0111 | 0.0011 | 0.0150 |

**Yelp**

| Core | BiasedMF | ImplicitMF | PopScore | User-KNN | Item-KNN | BPR | DiffRec | DMF | MultiVAE | Random | SimpleX |
|---|---|---|---|---|---|---|---|---|---|---|---|
| 0 | 0.0001 | 0.0077 | 0.0018 | 0.0079 | 0.0018 | 0.0040 | 0.0033 | 0.0017 | 0.0044 | 0.0000 | 0.0022 |
| 5 | 0.0002 | 0.0118 | 0.0022 | 0.0111 | 0.0038 | 0.0057 | 0.0083 | 0.0019 | 0.0022 | 0.0000 | 0.0032 |
| 10 | 0.0003 | 0.0208 | 0.0034 | 0.0188 | 0.0080 | 0.0098 | 0.0113 | 0.0026 | 0.0034 | 0.0000 | 0.0059 |
| 20 | 0.0005 | 0.0348 | 0.0054 | 0.0318 | 0.0164 | 0.0178 | 0.0152 | 0.0042 | 0.0028 | 0.0001 | 0.0106 |
| 50 | 0.0010 | 0.0569 | 0.0093 | 0.0561 | 0.0208 | 0.0238 | 0.0039 | 0.0081 | 0.0114 | 0.0003 | 0.0192 |
| 100 | 0.0013 | 0.0708 | 0.0152 | 0.0739 | 0.0155 | 0.0020 | 0.0003 | 0.0129 | 0.0019 | 0.0004 | 0.0059 |

**PRECISION@10 (PHASE 2)**

**Amazon CD**

| Core | BiasedMF | ImplicitMF | PopScore | User-KNN | Item-KNN | BPR | DiffRec | DMF | MultiVAE | Random | SimpleX |
|---|---|---|---|---|---|---|---|---|---|---|---|
| 5 | 0.0000 | 0.0005 | 0.0001 | 0.0009 | 0.0004 | 0.0170 | 0.0026 | 0.0010 | 0.0014 | 0.0000 | 0.0977 |
| 10 | 0.0000 | 0.0007 | 0.0002 | 0.0013 | 0.0006 | 0.0165 | 0.0025 | 0.0017 | 0.0013 | 0.0000 | 0.0993 |
| 20 | 0.0000 | 0.0010 | 0.0002 | 0.0018 | 0.0010 | 0.0142 | 0.0019 | 0.0009 | 0.0010 | 0.0000 | 0.0830 |
| 50 | 0.0000 | 0.0013 | 0.0004 | 0.0023 | 0.0012 | 0.0123 | 0.0008 | 0.0013 | 0.0007 | 0.0000 | 0.0476 |
| 100 | 0.0000 | 0.0019 | 0.0003 | 0.0024 | 0.0016 | 0.0102 | 0.0004 | 0.0010 | 0.0004 | 0.0000 | 0.0243 |

**Amazon Toys**

| Core | BiasedMF | ImplicitMF | PopScore | User-KNN | Item-KNN | BPR | DiffRec | DMF | MultiVAE | Random | SimpleX |
|---|---|---|---|---|---|---|---|---|---|---|---|
| 5 | 0.0000 | 0.0001 | 0.0000 | 0.0005 | 0.0002 | 0.1943 | 0.0005 | 0.0009 | 0.0000 | 0.0000 | 0.4327 |
| 10 | 0.0000 | 0.0001 | 0.0001 | 0.0005 | 0.0004 | 0.1387 | 0.0002 | 0.0009 | 0.0000 | 0.0000 | 0.2680 |
| 20 | 0.0000 | 0.0002 | 0.0000 | 0.0005 | 0.0003 | 0.0837 | 0.0001 | 0.0009 | 0.0000 | 0.0000 | 0.1384 |
| 50 | 0.0000 | 0.0002 | 0.0001 | 0.0004 | 0.0002 | 0.0343 | 0.0001 | 0.0005 | 0.0000 | 0.0000 | 0.0464 |
| 100 | 0.0000 | 0.0002 | 0.0000 | 0.0003 | 0.0003 | 0.0156 | 0.0000 | 0.0005 | 0.0000 | 0.0001 | 0.0172 |

**Gowalla**

| Core | BiasedMF | ImplicitMF | PopScore | User-KNN | Item-KNN | BPR | DiffRec | DMF | MultiVAE | Random | SimpleX |
|---|---|---|---|---|---|---|---|---|---|---|---|
| 5 | 0.0000 | 0.0011 | 0.0003 | 0.0019 | 0.0009 | 0.0136 | 0.0071 | 0.0034 | 0.0038 | 0.0000 | 0.1144 |
| 10 | 0.0000 | 0.0012 | 0.0003 | 0.0021 | 0.0010 | 0.0132 | 0.0069 | 0.0022 | 0.0039 | 0.0000 | 0.1189 |
| 20 | 0.0000 | 0.0017 | 0.0004 | 0.0027 | 0.0014 | 0.0132 | 0.0053 | 0.0047 | 0.0042 | 0.0000 | 0.1104 |
| 50 | 0.0000 | 0.0023 | 0.0005 | 0.0035 | 0.0017 | 0.0083 | 0.0033 | 0.0032 | 0.0029 | 0.0000 | 0.0730 |
| 100 | 0.0000 | 0.0025 | 0.0004 | 0.0034 | 0.0017 | 0.0055 | 0.0013 | 0.0018 | 0.0017 | 0.0000 | 0.0348 |

**MovieLens**

| Core | BiasedMF | ImplicitMF | PopScore | User-KNN | Item-KNN | BPR | DiffRec | DMF | MultiVAE | Random | SimpleX |
|---|---|---|---|---|---|---|---|---|---|---|---|
| 5 | 0.0002 | 0.0010 | 0.0009 | 0.0011 | 0.0011 | 0.0142 | 0.1479 | 0.0127 | 0.0151 | 0.0001 | 0.0109 |
| 10 | 0.0002 | 0.0011 | 0.0010 | 0.0011 | 0.0012 | 0.0130 | 0.1156 | 0.0132 | 0.0141 | 0.0001 | 0.0084 |
| 20 | 0.0005 | 0.0026 | 0.0026 | 0.0026 | 0.0030 | 0.0055 | 0.0289 | 0.0059 | 0.0096 | 0.0001 | 0.0027 |
| 50 | 0.0002 | 0.0009 | 0.0009 | 0.0008 | 0.0007 | 0.0034 | 0.0054 | 0.0059 | 0.0047 | 0.0001 | 0.0021 |
| 100 | 0.0003 | 0.0004 | 0.0003 | 0.0006 | 0.0003 | 0.0011 | 0.0020 | 0.0038 | 0.0034 | 0.0001 | 0.0004 |

**Yelp**

| Core | BiasedMF | ImplicitMF | PopScore | User-KNN | Item-KNN | BPR | DiffRec | DMF | MultiVAE | Random | SimpleX |
|---|---|---|---|---|---|---|---|---|---|---|---|
| 5 | 0.0000 | 0.0008 | 0.0002 | 0.0007 | 0.0003 | 0.0056 | 0.0218 | 0.0018 | 0.0023 | 0.0000 | 0.0070 |
| 10 | 0.0000 | 0.0010 | 0.0002 | 0.0010 | 0.0005 | 0.0050 | 0.0094 | 0.0013 | 0.0017 | 0.0000 | 0.0057 |
| 20 | 0.0000 | 0.0013 | 0.0002 | 0.0013 | 0.0007 | 0.0036 | 0.0046 | 0.0011 | 0.0017 | 0.0000 | 0.0049 |
| 50 | 0.0000 | 0.0012 | 0.0003 | 0.0011 | 0.0004 | 0.0024 | 0.0013 | 0.0008 | 0.0012 | 0.0000 | 0.0039 |
| 100 | 0.0000 | 0.0009 | 0.0003 | 0.0011 | 0.0003 | 0.0032 | 0.0005 | 0.0008 | 0.0008 | 0.0000 | 0.0018 |

**RECALL@10 (PHASE 1)**

**Amazon CD**

| Core | BiasedMF | ImplicitMF | PopScore | User-KNN | Item-KNN | BPR | DiffRec | DMF | MultiVAE | Random | SimpleX |
|---|---|---|---|---|---|---|---|---|---|---|---|
| 0 | 0.0000 | 0.0281 | 0.0096 | 0.0438 | 0.0126 | 0.0064 | 0.0111 | 0.0030 | 0.0095 | 0.0000 | 0.0038 |
| 5 | 0.0001 | 0.0262 | 0.0069 | 0.0428 | 0.0147 | 0.0053 | 0.0070 | 0.0027 | 0.0061 | 0.0000 | 0.0034 |
| 10 | 0.0001 | 0.0240 | 0.0059 | 0.0407 | 0.0190 | 0.0050 | 0.0049 | 0.0035 | 0.0039 | 0.0000 | 0.0025 |
| 20 | 0.0001 | 0.0249 | 0.0062 | 0.0435 | 0.0244 | 0.0053 | 0.0037 | 0.0017 | 0.0031 | 0.0000 | 0.0032 |
| 50 | 0.0009 | 0.0359 | 0.0085 | 0.0546 | 0.0308 | 0.0055 | 0.0004 | 0.0030 | 0.0024 | 0.0000 | 0.0039 |
| 100 | 0.0012 | 0.0518 | 0.0131 | 0.0741 | 0.0391 | 0.0060 | 0.0003 | 0.0032 | 0.0012 | 0.0000 | 0.0039 |

**Amazon Toys**

| Core | BiasedMF | ImplicitMF | PopScore | User-KNN | Item-KNN | BPR | DiffRec | DMF | MultiVAE | Random | SimpleX |
|---|---|---|---|---|---|---|---|---|---|---|---|
| 0 | 0.0000 | 0.0029 | 0.0010 | 0.0112 | 0.0040 | 0.0003 | 0.0000 | 0.0009 | 0.0005 | 0.0000 | 0.0005 |
| 5 | 0.0000 | 0.0022 | 0.0009 | 0.0106 | 0.0047 | 0.0003 | 0.0005 | 0.0008 | 0.0006 | 0.0000 | 0.0005 |
| 10 | 0.0000 | 0.0027 | 0.0011 | 0.0099 | 0.0049 | 0.0004 | 0.0005 | 0.0010 | 0.0007 | 0.0000 | 0.0005 |
| 20 | 0.0000 | 0.0029 | 0.0009 | 0.0092 | 0.0056 | 0.0002 | 0.0004 | 0.0007 | 0.0001 | 0.0000 | 0.0003 |
| 50 | 0.0000 | 0.0040 | 0.0012 | 0.0088 | 0.0057 | 0.0003 | 0.0003 | 0.0005 | 0.0002 | 0.0000 | 0.0003 |
| 100 | 0.0000 | 0.0065 | 0.0017 | 0.0094 | 0.0064 | 0.0005 | 0.0000 | 0.0003 | 0.0001 | 0.0000 | 0.0003 |

**Gowalla**

| Core | BiasedMF | ImplicitMF | PopScore | User-KNN | Item-KNN | BPR | DiffRec | DMF | MultiVAE | Random | SimpleX |
|---|---|---|---|---|---|---|---|---|---|---|---|
| 0 | 0.0003 | 0.0326 | 0.0072 | 0.0604 | 0.0254 | 0.0141 | 0.0089 | 0.0053 | 0.0062 | 0.0000 | 0.0077 |
| 5 | 0.0003 | 0.0333 | 0.0073 | 0.0601 | 0.0259 | 0.0138 | 0.0093 | 0.0038 | 0.0063 | 0.0000 | 0.0085 |
| 10 | 0.0003 | 0.0345 | 0.0077 | 0.0594 | 0.0271 | 0.0150 | 0.0077 | 0.0029 | 0.0063 | 0.0000 | 0.0070 |
| 20 | 0.0004 | 0.0368 | 0.0081 | 0.0591 | 0.0293 | 0.0141 | 0.0050 | 0.0028 | 0.0050 | 0.0000 | 0.0069 |
| 50 | 0.0007 | 0.0478 | 0.0106 | 0.0693 | 0.0349 | 0.0120 | 0.0005 | 0.0051 | 0.0017 | 0.0000 | 0.0046 |
| 100 | 0.0013 | 0.0695 | 0.0141 | 0.0894 | 0.0400 | 0.0093 | 0.0004 | 0.0035 | 0.0032 | 0.0000 | 0.0042 |

**MovieLens**

| Core | BiasedMF | ImplicitMF | PopScore | User-KNN | Item-KNN | BPR | DiffRec | DMF | MultiVAE | Random | SimpleX |
|---|---|---|---|---|---|---|---|---|---|---|---|
| 0 | 0.0089 | 0.0698 | 0.0639 | 0.0839 | 0.0736 | 0.0626 | 0.0565 | 0.0575 | 0.0668 | 0.0003 | 0.0405 |
| 5 | 0.0078 | 0.0606 | 0.0560 | 0.0702 | 0.0709 | 0.0561 | 0.0639 | 0.0469 | 0.0550 | 0.0003 | 0.0312 |
| 10 | 0.0061 | 0.0451 | 0.0426 | 0.0498 | 0.0542 | 0.0408 | 0.0503 | 0.0368 | 0.0411 | 0.0003 | 0.0239 |
| 20 | 0.0054 | 0.0322 | 0.0318 | 0.0332 | 0.0379 | 0.0294 | 0.0333 | 0.0242 | 0.0293 | 0.0003 | 0.0190 |
| 50 | 0.0056 | 0.0229 | 0.0245 | 0.0236 | 0.0244 | 0.0161 | 0.0140 | 0.0131 | 0.0063 | 0.0002 | 0.0126 |
| 100 | 0.0075 | 0.0219 | 0.0270 | 0.0249 | 0.0144 | 0.0049 | 0.0019 | 0.0099 | 0.0040 | 0.0004 | 0.0054 |

**Yelp**

| Core | BiasedMF | ImplicitMF | PopScore | User-KNN | Item-KNN | BPR | DiffRec | DMF | MultiVAE | Random | SimpleX |
|---|---|---|---|---|---|---|---|---|---|---|---|
| 0 | 0.0008 | 0.0449 | 0.0123 | 0.0481 | 0.0104 | 0.0229 | 0.0179 | 0.0112 | 0.0256 | 0.0001 | 0.0116 |
| 5 | 0.0007 | 0.0497 | 0.0110 | 0.0475 | 0.0157 | 0.0233 | 0.0321 | 0.0091 | 0.0109 | 0.0001 | 0.0127 |
| 10 | 0.0007 | 0.0493 | 0.0085 | 0.0432 | 0.0196 | 0.0205 | 0.0228 | 0.0061 | 0.0080 | 0.0001 | 0.0128 |
| 20 | 0.0007 | 0.0489 | 0.0080 | 0.0440 | 0.0250 | 0.0205 | 0.0172 | 0.0051 | 0.0031 | 0.0001 | 0.0125 |
| 50 | 0.0010 | 0.0569 | 0.0093 | 0.0561 | 0.0208 | 0.0121 | 0.0021 | 0.0043 | 0.0059 | 0.0002 | 0.0101 |
| 100 | 0.0013 | 0.0708 | 0.0152 | 0.0739 | 0.0155 | 0.0005 | 0.0001 | 0.0037 | 0.0007 | 0.0001 | 0.0015 |

**RECALL@10 (PHASE 2)**

**Amazon CD**

| Core | BiasedMF | ImplicitMF | PopScore | User-KNN | Item-KNN | BPR | DiffRec | DMF | MultiVAE | Random | SimpleX |
|---|---|---|---|---|---|---|---|---|---|---|---|
| 5 | 0.0000 | 0.0013 | 0.0003 | 0.0022 | 0.0009 | 0.0547 | 0.0073 | 0.0028 | 0.0043 | 0.0000 | 0.2751 |
| 10 | 0.0000 | 0.0012 | 0.0003 | 0.0021 | 0.0011 | 0.0326 | 0.0061 | 0.0038 | 0.0022 | 0.0000 | 0.1898 |
| 20 | 0.0000 | 0.0013 | 0.0003 | 0.0023 | 0.0013 | 0.0165 | 0.0048 | 0.0020 | 0.0011 | 0.0000 | 0.1028 |
| 50 | 0.0000 | 0.0014 | 0.0005 | 0.0024 | 0.0013 | 0.0080 | 0.0015 | 0.0033 | 0.0004 | 0.0000 | 0.0344 |
| 100 | 0.0000 | 0.0019 | 0.0003 | 0.0024 | 0.0016 | 0.0048 | 0.0011 | 0.0038 | 0.0002 | 0.0000 | 0.0120 |

**Amazon Toys**

| Core | BiasedMF | ImplicitMF | PopScore | User-KNN | Item-KNN | BPR | DiffRec | DMF | MultiVAE | Random | SimpleX |
|---|---|---|---|---|---|---|---|---|---|---|---|
| 5 | 0.0000 | 0.0001 | 0.0000 | 0.0005 | 0.0002 | 0.1943 | 0.0005 | 0.0009 | 0.0000 | 0.0000 | 0.4327 |
| 10 | 0.0000 | 0.0001 | 0.0001 | 0.0005 | 0.0004 | 0.1387 | 0.0002 | 0.0009 | 0.0000 | 0.0000 | 0.2680 |
| 20 | 0.0000 | 0.0002 | 0.0000 | 0.0005 | 0.0003 | 0.0837 | 0.0001 | 0.0009 | 0.0000 | 0.0000 | 0.1384 |
| 50 | 0.0000 | 0.0002 | 0.0001 | 0.0004 | 0.0002 | 0.0343 | 0.0001 | 0.0005 | 0.0000 | 0.0000 | 0.0464 |
| 100 | 0.0000 | 0.0002 | 0.0000 | 0.0003 | 0.0003 | 0.0156 | 0.0000 | 0.0005 | 0.0000 | 0.0001 | 0.0172 |

**Gowalla**

| Core | BiasedMF | ImplicitMF | PopScore | User-KNN | Item-KNN | BPR | DiffRec | DMF | MultiVAE | Random | SimpleX |
|---|---|---|---|---|---|---|---|---|---|---|---|
| 5 | 0.0000 | 0.0018 | 0.0004 | 0.0034 | 0.0015 | 0.0224 | 0.0097 | 0.0050 | 0.0062 | 0.0000 | 0.2267 |
| 10 | 0.0000 | 0.0019 | 0.0004 | 0.0032 | 0.0016 | 0.0188 | 0.0086 | 0.0032 | 0.0060 | 0.0000 | 0.1930 |
| 20 | 0.0000 | 0.0021 | 0.0005 | 0.0034 | 0.0017 | 0.0134 | 0.0075 | 0.0066 | 0.0067 | 0.0000 | 0.1234 |
| 50 | 0.0000 | 0.0024 | 0.0005 | 0.0036 | 0.0018 | 0.0061 | 0.0050 | 0.0054 | 0.0053 | 0.0000 | 0.0494 |
| 100 | 0.0000 | 0.0025 | 0.0004 | 0.0034 | 0.0017 | 0.0032 | 0.0020 | 0.0055 | 0.0043 | 0.0000 | 0.0153 |

**MovieLens**

| Core | BiasedMF | ImplicitMF | PopScore | User-KNN | Item-KNN | BPR | DiffRec | DMF | MultiVAE | Random | SimpleX |
|---|---|---|---|---|---|---|---|---|---|---|---|
| 5 | 0.0004 | 0.0031 | 0.0028 | 0.0034 | 0.0034 | 0.0491 | 0.4559 | 0.0454 | 0.0554 | 0.0003 | 0.0352 |
| 10 | 0.0004 | 0.0024 | 0.0023 | 0.0024 | 0.0028 | 0.0368 | 0.2940 | 0.0436 | 0.0487 | 0.0003 | 0.0193 |
| 20 | 0.0010 | 0.0059 | 0.0060 | 0.0061 | 0.0070 | 0.0225 | 0.0854 | 0.0280 | 0.0544 | 0.0003 | 0.0058 |
| 50 | 0.0003 | 0.0011 | 0.0011 | 0.0010 | 0.0009 | 0.0122 | 0.0178 | 0.0347 | 0.0244 | 0.0003 | 0.0022 |
| 100 | 0.0003 | 0.0006 | 0.0004 | 0.0007 | 0.0003 | 0.0050 | 0.0119 | 0.0303 | 0.0260 | 0.0003 | 0.0005 |

**Yelp**

| Core | BiasedMF | ImplicitMF | PopScore | User-KNN | Item-KNN | BPR | DiffRec | DMF | MultiVAE | Random | SimpleX |
|---|---|---|---|---|---|---|---|---|---|---|---|
| 5 | 0.0000 | 0.0025 | 0.0006 | 0.0024 | 0.0008 | 0.0213 | 0.0941 | 0.0082 | 0.0112 | 0.0001 | 0.0303 |
| 10 | 0.0000 | 0.0026 | 0.0005 | 0.0022 | 0.0011 | 0.0149 | 0.0299 | 0.0058 | 0.0083 | 0.0001 | 0.0163 |
| 20 | 0.0000 | 0.0026 | 0.0005 | 0.0025 | 0.0014 | 0.0081 | 0.0102 | 0.0055 | 0.0087 | 0.0001 | 0.0094 |
| 50 | 0.0000 | 0.0016 | 0.0004 | 0.0017 | 0.0006 | 0.0046 | 0.0031 | 0.0049 | 0.0083 | 0.0001 | 0.0055 |
| 100 | 0.0000 | 0.0015 | 0.0004 | 0.0014 | 0.0006 | 0.0046 | 0.0021 | 0.0060 | 0.0064 | 0.0001 | 0.0026 |